\documentclass[conference]{IEEEtran}

\IEEEoverridecommandlockouts  
\makeatletter
\def\@IEEEpubidpullup{} 
\def\@IEEEpubid#1{}     
\makeatother

\ifCLASSINFOpdf
\else
\fi

\usepackage{hyperref}
\usepackage[pdftex]{graphicx}
\usepackage{xcolor}
\usepackage{enumitem}
\usepackage{listings}
\usepackage{array}
\usepackage{amsmath}

\usepackage{times}
\usepackage{latexsym}

\usepackage[T1]{fontenc}

\usepackage[utf8]{inputenc}

\usepackage{microtype}

\usepackage{inconsolata}

\usepackage{graphicx}
\usepackage{multirow}
\usepackage{float}
\usepackage{amsmath}
\usepackage{dsfont}

\usepackage{xcolor, soul} 
\usepackage[x11names]{xcolor}

\usepackage{bm}
\usepackage{amssymb}
\usepackage{enumitem}
\usepackage{tabularx}
\usepackage{booktabs}
\usepackage{array}  
\usepackage[absolute,overlay]{textpos}
\usepackage{algorithm}
\usepackage{algorithmic}

\usepackage{stfloats}

\usepackage{amsfonts}

\definecolor{MyDarkGreen}{RGB}{1, 69, 35} 
\definecolor{MyDarkGray}{RGB}{41, 41, 40} 

\newlist{tableitemize}{itemize}{1}
\setlist[tableitemize,1]{noitemsep, topsep=0pt, left=0pt, label=--}

\newif\ifanonymous
\anonymousfalse  

\newcommand{\IfAnon}[2]{\ifanonymous #1\else #2\fi} 

\newcommand{\AnonText}[1]{\IfAnon{#1}{}}               
\newcommand{\RealText}[1]{\IfAnon{}{#1}}               
\newcommand{\SwapText}[2]{\IfAnon{#1}{#2}}             

\newcommand{\selfcite}[2]{\IfAnon{#1}{#2}}

\newcommand{\blindurl}[1]{\IfAnon{\texttt{[link hidden for review]}}{\url{#1}}}
\newcommand{\blindemail}[1]{\IfAnon{\texttt{[email hidden]}}{\texttt{#1}}}

\newif\ifieee
\ieeefalse   

\usepackage{pdfpages}

\begin{document}

\ifanonymous
\linenumbers
\else
\fi

\title{From Measurement to Expertise: Empathetic Expert Adapters for Context-Based Empathy in Conversational AI Agents}


\ifanonymous
\author{Anonymous Authors}
\else
\author{%
  Erfan Shayegani\IEEEauthorrefmark{1}\IEEEauthorrefmark{2},
  Jina Suh\IEEEauthorrefmark{1},
  Andy Wilson\IEEEauthorrefmark{1},
  Nagu Rangan\IEEEauthorrefmark{1},
  Javier Hernandez\IEEEauthorrefmark{1}
  
  \thanks{\IEEEauthorrefmark{1}Microsoft Research. Corresponding authors: \texttt{javierh@microsoft.com}, \texttt{sshay004@ucr.edu}}%
  \thanks{\IEEEauthorrefmark{2}University of California, Riverside. \texttt{work performed while at Microsoft Research.}}%
}
\fi

\markboth{Journal of \LaTeX\ Class Files,~Vol.~14, No.~8, August~2015}%
{Shell \MakeLowercase{\textit{et al.}}: Bare Demo of IEEEtran.cls for IEEE Journals}



\maketitle

\begin{abstract}


Empathy is a critical factor in fostering positive user experiences in conversational AI. While models can display empathy, it is often generic rather than tailored to specific tasks and contexts. 
In this work, we introduce a novel framework for developing and evaluating context-specific empathetic large language models (LLMs). We first analyze a real-world conversational dataset consisting of 672 multi-turn conversations across 8 tasks, revealing significant differences in terms of expected and experienced empathy before and after the conversations, respectively. 
To help minimize this gap, we develop a synthetic multi-turn conversational generation pipeline and steer responses toward our defined empathy patterns based on the context that more closely matches users' expectations. We then train empathetic expert adapters for context-specific empathy that specialize in varying empathy levels based on the recognized task. Our empirical results demonstrate a significant gap reduction of 72.66\% between perceived and desired empathy
with scores increasing by an average factor of 2.43 as measured by our metrics and reward models. Additionally, our trained empathetic expert adapters demonstrate superior effectiveness in preserving empathy patterns throughout conversation turns, outperforming system prompts, which tend to dramatically diminish in impact as conversations lengthen.


\end{abstract}



%
\IEEEpeerreviewmaketitle



\section{Introduction}

Empathy is a cornerstone of meaningful human interactions and plays a crucial role in the quality of conversations, particularly when dealing with sensitive and personal topics. As large language models (LLMs) are increasingly integrated into everyday applications, their ability to effectively exhibit empathy becomes paramount. AI agents are being deployed across a wide variety of contexts—from programming assistance to mental health support—that require varying levels of empathy. However, existing conversational AI systems struggle to demonstrate context-appropriate empathy and tend to provide uniform empathy levels, leading to unsatisfactory user experiences. This challenge is even more pronounced when empathy needs to be tailored based on specific user contexts or tasks, such as providing emotional support during a distressing life event versus offering guidance for professional development.

Several studies have explored empathy in conversational AI, with many relying on the inherent reasoning and commonsense abilities of LLMs to produce generic empathetic responses, often lacking task-specific focus and contextual adaptability. Most of these works are prompt-based, investigating the impact of techniques like Chain-of-Thought (CoT), Retrieval-Augmented Generation (RAG), self-reflection, and similar prompting methods to enhance empathetic expression \cite{chen2024causeawareempatheticresponsegenerationEE1, lee2023chainEE1, VeraSorin2023LargeLMSURVEYEE1, chen2023soulchatimprovingllmsempathyEE1, croissant2023appraisalbasedchainofemotionarchitectureaffectiveEE1, sotolar2024empoemotiongroundingempatheticDPOEE1, qian2024harnessingpowerlargelanguageEE1, lee2024largelanguagemodelsproduceJINAEE1, sabour2021cemcommonsenseawareempatheticresponseCOMMEE1,rashkin2019empatheticopendomainconversationmodelsEDEE1, liang2024synthempathyhighqualitysyntheticempathyEE1}. Several other studies are focused on empathy in narrow domains like healthcare, psychotherapy, doctor-patient interactions, mental health counseling, group identities, and emotional intelligence, which restricts their adaptability and generalizability, rendering them less applicable for diverse empathy needs across broader contexts \cite{lee2023chainEE1,chen2024emotionqueenbenchmarkevaluatingempathyQueenEE2, wang2024giebenchholisticevaluationgroupEE2, huang2024emotionallynumbempatheticevaluatingEE2, loh2023harnessinglargelanguagemodelsEE2, luo2024assessingempathylargelanguageEE2, sabour2024emobenchevaluatingemotionalintelligenceEE2}.

Motivated by these limitations, in this work, we introduce a novel framework for developing and evaluating contextually empathetic LLMs. We start by analyzing a real-world conversational dataset~\cite{suh2025Sense7} that includes users' pre-desired empathy levels, post-conversation perceived empathy ratings, and qualitative feedback on interactions with AI assistants. These insights reveal that empathy is highly context-dependent and that the perceived quality of conversations is significantly correlated with the level and type of empathy demonstrated by the assistant. We leverage these findings to inform the design of empathy measurement methods and reward models that evaluate and refine the empathetic capabilities of LLMs.

To address the limitations of data scarcity and variability in empathy-driven conversations, we develop a synthetic multi-turn conversation generation pipeline. By clustering tasks with similar empathy requirements, we generate diverse empathetic conversations, steering responses toward well-defined empathy patterns based on real-world user insights. To further enhance empathy modeling, we employ empathetic expert adapters trained specifically for context-sensitive empathy, ensuring that the LLM is capable of adapting its empathetic tone and style across a range of user needs.

Our evaluation spans multiple experimental setups, including a Baseline LLM without empathy-specific training used in its inherent form, a System Prompt approach where empathy guidelines are injected into the context of the baseline LLM, and our proposed Empathetic Expert Adapter approach. Empirical results demonstrate that the Empathetic Expert Adapters outperform both the Baseline and System Prompt settings, consistently aligning with the defined empathy patterns and maintaining empathy across multi-turn conversations. The ability of Empathetic Expert Adapters to sustain empathy throughout extended dialogues makes them a powerful tool for improving user satisfaction in conversational AI systems.

This study not only showcases the potential of training contextually empathetic agents but also introduces practical methodologies for empathy measurement, synthetic data augmentation, and context-specific fine-tuning. By advancing the empathetic abilities of LLMs, we take an important step towards creating conversational agents that can more effectively adapt to the users' needs, enhancing their experience in emotionally meaningful and impactful ways.

The paper is organized as follows: Section~\ref{sec:background} provides a review of related work, highlighting the state of empathy modeling in AI and its limitations. Section~\ref{sec:methodology} describes our methodology, detailing the analysis of real-world conversations, synthetic data generation, and the training of empathetic expert adapters. Section~\ref{sec:experiments} presents our experimental setup and evaluation results, showcasing the effectiveness of our approach across various task contexts. Section~\ref{sec:discuss} discusses the implications of our findings, comparing the performance of different empathy modeling approaches and the impact of underlying architectures. Finally, Section~\ref{sec:conclusion} concludes the paper with a summary of key contributions, limitations, and future research directions.

\begin{figure*}[t]
    \begin{center} 
        \includegraphics[width=0.75\linewidth]{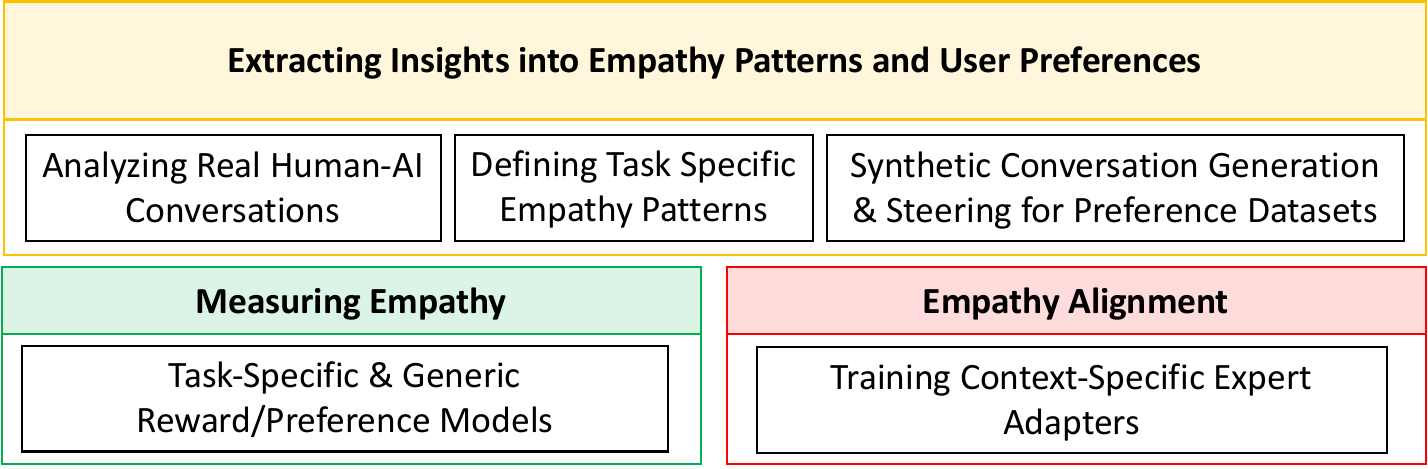}
    \end{center}
    \caption{Our approach consists of multiple stages: extracting insights from real human-AI interactions, defining task-specific empathy patterns, generating synthetic conversations, and steering them for preference datasets. We then measure empathy using task-specific and generic reward models followed by an alignment stage where context-specific empathetic expert adapters are trained to enhance empathetic responses.}
    \label{fig:MethodStages}
\end{figure*}


\section{Background}\label{sec:background}



Empathy in AI agents has been a focal point of active research, particularly in the development of chatbots and social robots that aim to support and engage users more effectively. Empathetic AI agents are designed to recognize and respond to human emotions and experiences, thereby fostering more natural and meaningful interactions. For instance, Leite et al.~\cite{leite2013influence} emphasized the role of empathy in human–robot interaction, suggesting that empathetic behaviors can significantly enhance user engagement. Similarly, Paiva et al.~\cite{paiva2017emotion} provided a comprehensive overview of emotion modeling for social robots, highlighting the potential of empathetic agents in various applications. Other studies have also explored the integration of empathy in conversational agents to improve user satisfaction and trust. For example, Lubis et al.~\cite{lubis2018eliciting} developed an empathetic dialog system that elicits positive emotions through affective backchannels.

To facilitate research in empathetic AI agents, several studies have been conducted to collect relevant datasets. For instance, Rashkin et al.~\cite{rashkin2019empatheticopendomainconversationmodelsEDEE1} introduced the ``EmpatheticDialogues'' dataset of 25K conversations between two people grounded in emotional situations, which has been instrumental in training models for empathetic response generation. In a separate effort, Liu et al.~\cite{liu2021towards} introduced the ``Emotional Support Conversation'' dataset, consisting of 10K multi-turn dialogues focused on providing emotional support, aiming to model how humans offer assistance in distressing situations. While these and other datasets have enabled much research in the context of empathetic AI agents, they do not fully capture the nuances of context-specific empathy required in diverse interaction scenarios. 
To address this gap, our work builds on the recently released \emph{SENSE-7} dataset by Suh et al.~\cite{suh2025Sense7}, which contains 672 conversations where 109 participants provided per-turn subjective ratings of empathy while interacting with AI agents. Using this dataset, we conduct a detailed analysis that expands on their findings and provides deeper insights into how people’s expectations and perceptions of empathy vary across different interaction contexts.

Studies on context-specific empathetic agents have typically focused on exploring one application at a time, tailoring the agent's empathetic responses to a specific domain. For instance, Lisetti et al.~\cite{lisetti2013i} developed an empathetic virtual counselor to promote behavior change in healthcare settings, while Fitzpatrick et al.~\cite{fitzpatrick2017delivering} evaluated a conversational agent named Woebot for delivering cognitive behavioral therapy to young adults with symptoms of depression and anxiety. These agents demonstrate the effectiveness of empathy in specialized contexts but are limited in their ability to adapt to different user tasks or scenarios. In contrast, our work aims to understand the user's context dynamically and adjust the agent's empathetic displays accordingly, providing a more versatile and naturalistic interaction experience across various applications.



\section{Methodology}\label{sec:methodology}

Our approach consists of multiple stages, beginning with extracting insights from our real-world conversational dataset, followed by the development of synthetic data generation, steering, reward models, and training empathetic expert adapters for context-specific empathy.

\subsection{Insights from Real Conversations} 
We extend the analysis of a real-world conversational dataset (\emph{SENSE-7}~\cite{suh2025Sense7}), which includes 672 multi-turn interactions between users and AI assistants across eight distinct tasks conducted ``in the wild'' with participants interacting naturally with the chatbots who reported their desired and perceived level of empathy before and after each conversation, respectively. In particular, participants reported perceived empathy along multiple dimensions such as Affective, Cognitive, and Contextual. The eight tasks include navigating work issues, career/self-improvement challenges, overcoming personal issues, writing assistance, thinking through a work assignment, learning new skills, handling distressing situations, and navigating social situations.
Participants also responded to several 5-point Likert scale questions focused on assessing the level at which the AI assistant achieved success, the level of user engagement, the positivity of the interaction, and the likelihood of future use among others. In addition, participants also provided qualitative feedback on their experience. The dataset includes 151 participants, with 109 completing at least five conversations and the exit survey, resulting in 672 fully labeled conversations. 
Our analysis of the dataset yielded two primary findings with further details included in the Appendix A:

- Average perceived empathy highly correlates with self-reported satisfaction after the conversation (Pearson's Correlation = 0.64).

- Desired empathy varies significantly depending on the task and context. For example, as shown by the pre-desired empathy bars in Figure~\ref{fig:final_open}, tasks around overcoming personal issues and distressing situations require much more empathy compared to tasks on learning new skills or thinking through work assignments. 

These insights were crucial for shaping the empathy measurement models and guiding our synthetic data generation and training processes.


\subsection{Synthetic Conversational Pipeline} \label{subsec:synthetic}

Based on our analysis of the real-world dataset, we observe consistent patterns in the relationship between the gap in pre-desired empathy levels and the expected post-perceived empathy, as well as user satisfaction signals (Appendix B-B). Leveraging these insights, we group the eight tasks into four distinct clusters—\textit{Distressing/Social/Personal Situations}, \textit{Learning Skills}, \textit{Work Issues/Career/Self-Improvement}, and \textit{Work Assignment/Help with Writing}—each sharing similar levels and types of empathy; thereby streamlining the process of generating new synthetic conversations as well as designing and steering the responses based on the shared empathy patterns within each group. 

\paragraph{Multi-Turn Conversation Generation} 

Our multi-turn generation pipeline consists of several stages. We first leverage GPT-4o to come up with initial questions per task cluster, simulating the opening prompts of the users. To ensure proximity to our real dataset distribution, we design detailed prompts consisting of In-Context Learning (ICL) examples of our real dataset. Figure 10 shows the prompt used with GPT-4o for the generation of the first user turns mimicking the initial prompt at the start of a conversation. 

Figure 11 shows a sample of the corresponding generated questions. In order to be close to the original distribution of our real dataset, we incorporate several In-Context Learning (ICL) examples in the body of the prompt as well as a description of the task category at the beginning of the prompt which in Figure 10 is the ``Distressing/Social/Personal Situations'' task cluster. 

To maximize the diversity of the generated questions and to avoid generic questions, we ask GPT-4o to produce 30 different questions and call the LLM 40 times with varying decoding parameters (Temperature and Top-p). We also expand the exploratory space of the LLM by asking it to: 

``\textit{When generating questions, consider those that individuals from diverse backgrounds, such as men and women, managers, researchers, CEOs, parents, engineers, family members, and workers aged 20 to 65 might ask an AI assistant.
Don't ask general questions; be very specific and detailed in your questions, depicting real-life scenarios, situations, people, concepts, and emotions.}''
Next, we utilize Llama-3-8B-Instruct's~\cite{dubey2024LLAMA3Models} auto-regression capabilities to generate subsequent assistant and user turns. The assistant's response is generated directly from the initial user input. For user turns, we employ the Chat Template, appending ``\textit{<|start\_header\_id|>user<|end\_header\_id|>\textbackslash n\textbackslash n}'' to the previous context. The LLM's auto-regressive nature then predicts the next user tokens. This process continues iteratively until reaching the desired number of turns.

\begin{figure}[ht]
\centering

\colorbox{pink!35} 
{
    \parbox{7cm}
    {
        \textbf{System Prompt:}\\
        \textit{``You are an AI assistant designed to be helpful. The user will engage with you in a **multi-turn conversation** revolving around \{TASK\_CLUSTER\}, where they will ask questions that build upon your previous responses, just like in a typical chat. Each user prompt will be relevant to the prior exchanges, with follow-up questions conditioned on the previous context. Your objective is to offer comprehensive, relevant, and insightful answers to assist the user throughout the conversation''.}
    }
}

\caption{System prompt for multi-turn coherency. The TASK\_CLUSTER variable can be one of: \textit{Distressing/Social/Personal Situations}, \textit{Learning Skills}, \textit{Work Issues/Career/Self-Improvement}, and \textit{Work Assignment/Help with Writing}}
\label{fig:sys_prompt}
\end{figure}

In our experiments, we notice sometimes the LLM does not stick to the previous context and predicts the next user turn which is not coherent with the previous context leading to a topic drift. This, in fact, happens in real conversations as well, however, we prioritize maintaining coherence to align with our real dataset's conversation styles which are mostly focused on a single concept along the conversation. To achieve this, we supervise Llama-3-8B-Instruct with a system prompt simulating a multi-turn conversation as shown in Figure~\ref{fig:sys_prompt}. 

We observed that more complex and lengthy system prompts lead to degraded generation quality. While we initially considered incorporating more task-specific conversation examples in the prompts, this approach increased the likelihood of significant drifts and inconsistent follow-ups. Therefore, we stick to the concise system prompt format as shown in Figure~\ref{fig:sys_prompt}. 

We observed the same phenomenon with lengthy system prompts and the inability of the LLMs to adhere to them especially in multi-turn long chat scenarios as discussed in Section~\ref{sec:discuss}. Additional details and examples are presented in Appendix B-A. 

Algorithm~\ref{alg:synthetic_multiturn} outlines the details of our synthetic multi-turn conversation generation pipeline.

\begin{algorithm*}[t]
\caption{Synthetic Multi-turn Dialogue Generation Pipeline}
\label{alg:synthetic_multiturn}
\begin{algorithmic}[1]
\REQUIRE 
  \begin{itemize}
    \item $\mathcal{C} = \{c_1, c_2, \ldots, c_N\}$: Set of $N$ task clusters
    \item $\mathcal{D}$: Real dataset for In-Context Learning (ICL) examples
    \item $T$: Desired number of user--assistant turns (excluding the initial user prompt)
    \item \textsc{GPT-4o}: LLM used for initial user prompts
    \item \textsc{Llama-3-8B-Instruct}: LLM used for subsequent turns
  \end{itemize}
\ENSURE 
  \begin{itemize}
    \item $\mathcal{S}$: A set of synthetic multi-turn dialogues
  \end{itemize}

\STATE $\mathcal{S} \gets \{\}$ \quad \COMMENT{Initialize empty collection of dialogues}

\FOR{each task cluster $c$ in $\mathcal{C}$}
  \STATE \COMMENT{\textbf{Step 1: Generate the initial user prompt using GPT-4o}}
  \STATE $\text{prompt}_{\text{ICL}} \gets \text{BuildPromptICL}(\mathcal{D}, c)$
  \STATE $u_0 \gets \textsc{GPT-4o}(\text{prompt}_{\text{ICL}})$

  \STATE \COMMENT{\textbf{Step 2: Initialize the conversation context}}
  \STATE $\text{conversation} \gets [(\texttt{user}, u_0)]$

  \STATE \COMMENT{\textbf{Step 3: Iteratively generate assistant and user turns}}
  \FOR{$t = 1$ to $T$}
    \STATE $a_t \gets \textsc{Llama-3-8B-Instruct}(\text{conversation})$ \quad \COMMENT{Assistant response}
    \STATE \text{append } $(\texttt{assistant}, a_t)$ to $\text{conversation}$

    \STATE $\text{template} \gets \text{"\textbf{\textit{<|start\_header\_id|>user<|end\_header\_id|>\textbackslash n\textbackslash n"}}}$
    \STATE $u_t \gets \textsc{Llama-3-8B-Instruct}(\text{conversation} + \text{template})$ \quad \COMMENT{Next user turn}
    \STATE \text{append } $(\texttt{user}, u_t)$ to $\text{conversation}$
  \ENDFOR

  \STATE \COMMENT{\textbf{Store the fully generated conversation}}
  \STATE $\mathcal{S} \gets \mathcal{S} \cup \{\text{conversation}\}$
\ENDFOR

\RETURN $\mathcal{S}$

\end{algorithmic}
\end{algorithm*}

\paragraph{Steering Toward Empathy Preferences} 

Building on our earlier insights, we define specific empathy levels and patterns for each task cluster. Two of the authors of this paper, drawing on their backgrounds in empathy and affective sciences, Human-Computer Interaction (HCI), and computer science, collaborated to design the target empathy patterns, ensuring that responses within each task cluster align with the desired empathy goals.

Incorporating these defined empathy patterns, we craft prompts that guide the assistant's responses toward either empathetic or non-empathetic behaviors within each task cluster. We utilize GPT-4o, first presenting the entire conversation context, and then simulating the dialogue turn by turn, instructing the model to follow the empathy guidelines embedded in our prompts. Finally, we steer both the real and synthetic conversations. Please see Appendix B-B for more details.


\subsection{Reward Models} \label{sec:rewards}

To quantify and measure empathy, we develop models that score conversations on a continuous scale from 0 to 1. 
In particular, we focus on two well-known types of reward modeling approaches \cite{lambert2024rewardbench}; Generative (LLM-as-a-judge) and Learning-based Classifiers. 

\paragraph{Generative}

We leverage GPT-4o~\cite{openai2023gpt4o} as a judge to analyze the conversation history between users and AI assistants, assigning scores based on the assistant's empathy alignment with our pre-defined empathy patterns. We incorporate different prompt engineering approaches~\cite{yang2024LLMOptimizer,wang2024mixtureMoA,chen2024LLMcallScale} which we refer to in Appendix C-A.


\paragraph{Learning-based Classifiers}

We attach a regression head $w\in \mathbb{R}^{d}$ on top of an LLM backbone (fsFairX-LLaMA3-RM-v0.1) as a complex feature extractor $f_\theta$ without the original output linear layer, where ${d}$ is the hidden dimension of the backbone. The whole conversation ${x}$ from the dataset distribution ${D}$ is then fed to the backbone $f_\theta$ and we extract the embedding of the conversation $f_\theta(x)$ and feed it to the regression head ${w}$ which outputs the score prediction. 



\begin{align}\label{eq:regression}
\begin{split}
&\min_{\theta, w} \mathbb{E}_{x\in D} \Bigg[
    \|w^\top f_\theta(x) - r\|_2^2 \\
    &-\log \frac{\exp(\beta R_{\mathrm{+}})}{\exp(\beta R_\mathrm{+}) + \exp(\beta r) + \exp(\beta R_\mathrm{-})} \\
    &-\log \frac{\exp(\beta r)}{\exp(\beta r) + \exp(\beta R_\mathrm{-})}
\Bigg]
\end{split}
\end{align}

Having the ground truth scalar reward score ${r\in \mathbb{R}}$ from our real dataset, the model is trained with our loss combining a regression term and two Bradley-Terry (BT) terms~\cite{Bradley1952BT} as shown in Eqn.~\ref{eq:regression} with an additional scaling variable, ${\beta\in \mathbb{R}}$, while keeping the LLM backbone frozen~\cite{wang2024ArmoRM} and only training the regression head. ${R_\mathrm{+}} = f_\theta(x_\mathrm{+})$ and ${R_\mathrm{-}} = f_\theta(x_\mathrm{-})$ are the scores our reward model predicts for the steered conversations aligned with our defined empathetic and non-empathetic patterns. The regression loss supervises the model to produce reward predictions that closely approximate the ground truth values, while the BT terms give it a sense of preferences among different empathy patterns and levels in an unsupervised manner. Our second loss term teaches the model that the empathetic-steered conversations should get better scores than the original conversations and the non-empathetic ones. 
($f_\theta(x_\mathrm{+}) > f_\theta(x)$
AND $f_\theta(x_\mathrm{+}) > f_\theta(x_\mathrm{-})$); The third loss term teaches it to give a lower score to the non-empathetic steered conversations compared to the original conversations. 
($f_\theta(x) > f_\theta(x_\mathrm{-})$). 
We have two main objectives; alignment with our predefined empathy patterns, and alignment with users' preferences and their pre-desired empathy. For the former, we train task-specific reward models with the mentioned loss on our real dataset divided by task clusters which provide necessary ground truth scalar reward scores assigned by the participants. Additionally, we train task-specific BT-based preference models using a combination of real and synthetic conversations. For the latter, we also train a generic reward model eliminating the BT loss terms and only including the regression term on all the tasks combined. 

Table~\ref{tab:RewardModel_hyperparams} and Table~\ref{tab:prefModel_hyperparams} include the reward and preference models' training details. 

\begin{figure*}[ht]
    \begin{center} 
        \includegraphics[width=1\linewidth]{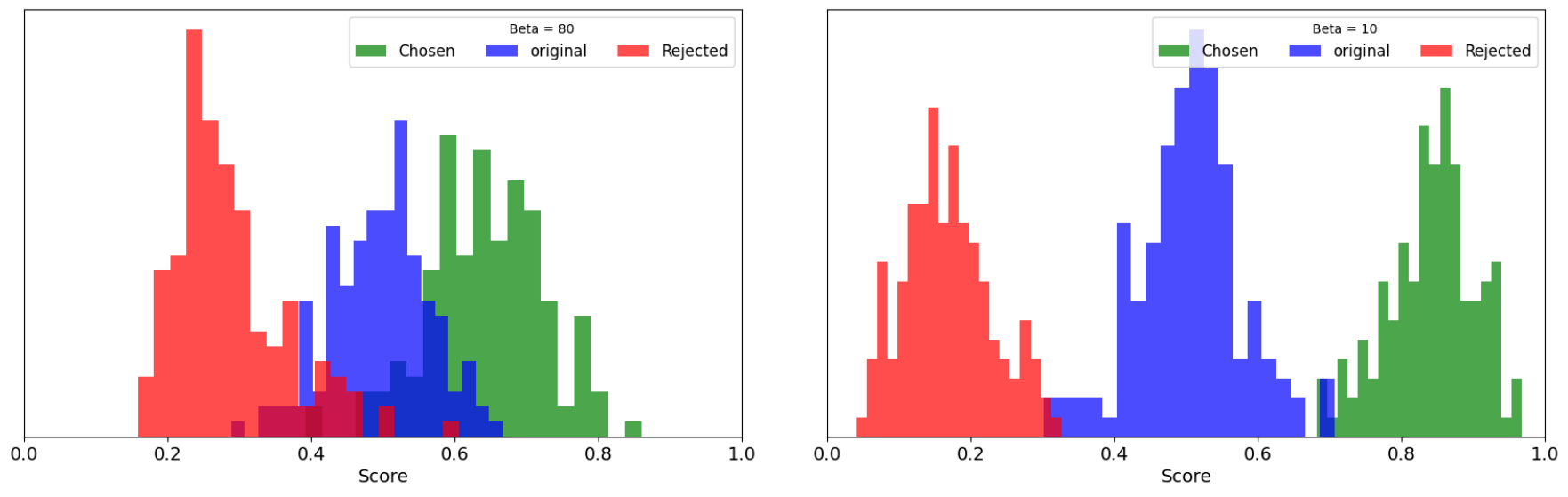}
    \end{center}
    \caption{The trained preference model's predictions. $\beta$ sets the sensitivity and the sharpness of the preference model. The smaller, the sharper. `Chosen' corresponds to the empathetic steered conversations, while `Rejected' means the non-empathetic steered conversations. The preference model has successfully learned to assign higher scores to our defined empathy patterns, and lower scores to the non-empathetic conversations.}
    \label{fig:pref_distrib}
\end{figure*}

\renewcommand{\arraystretch}{1.3}
\begin{table}[ht]
\centering
\resizebox{1\linewidth}{!}{
\begin{tabular}{cc|cc}
\toprule
Parameters            & Values & Parameters     & Values         \\ \toprule \toprule
Per-Device Batch size          & 16      & Learning rate  & 1e-6         \\
Number of Epochs  & 1000   & LLM Backbone   & fsFairX-LLaMA3-RM-v0.1          \\
Head Input dim  & 4096   & Head Middle dim   & 512          \\
Head Hidden dim  & 64   & Head Output dim   & 1          \\
Optimizer     & Adam      & Loss & Eqn.~\ref{eq:regression} \\ \bottomrule \bottomrule
\end{tabular}
}
\caption{Parameters for reward model training. The regression head is a simple Multi-Layer Perceptron (MLP) with the mentioned dimensions.}
\label{tab:RewardModel_hyperparams}
\end{table}

Further details can be found in Appendix C-B.

\begin{figure*}[ht]
    \begin{center} 
        \includegraphics[width=0.7\linewidth]{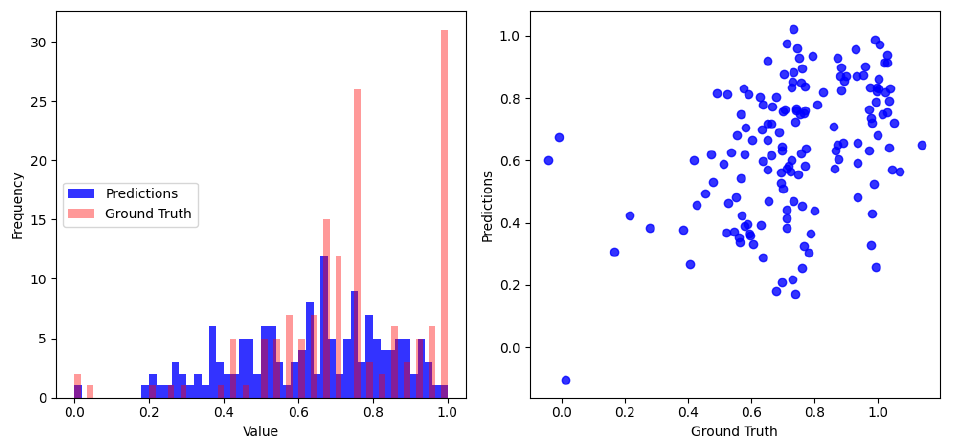}
    \end{center}
    \caption{The generic reward model's predictions. \textbf{MSE} = 0.0301, \textbf{MAE} = 0.1335, \textbf{Correlation} (Ground Truth, Predictions) = 0.43}
    \label{fig:reward_MSE}
\end{figure*}

\renewcommand{\arraystretch}{1}
\begin{table}[htbp]
\centering
\resizebox{0.8\linewidth}{!}{
\begin{tabular}{ll}
\toprule
Statistics & Values \\ \midrule \midrule
MSE & 0.0301 \\
MAE &  0.1335\\
Correlation (Ground Truth, Predictions) & 0.43 \\ \bottomrule \bottomrule

\end{tabular}
}
\vspace{1mm}
\caption{Generic Reward model evaluation.}
\label{tab:reward_MSE}
\end{table}

\renewcommand{\arraystretch}{1.3}
\begin{table}[ht]
\centering
\resizebox{1\linewidth}{!}{
\begin{tabular}{cc|cc}
\toprule
Parameters            & Values & Parameters     & Values         \\ \toprule \toprule
Per-Device Batch size          & 16      & Learning rate  & 5e-4         \\
Number of Epochs  & 150   & LLM Backbone   & fsFairX-LLaMA3-RM-v0.1          \\
Head Input dim  & 4096   & Head Middle dim   & 512          \\
Head Hidden dim  & 64   & Head Output dim   & 1          \\
Optimizer     & Adam      & Loss & Eqn. 2 \\ \bottomrule \bottomrule
\end{tabular}
}
\caption{Parameters for preference model (BT) training. The regression head is a simple Multi-Layer Perceptron (MLP) with the mentioned dimensions.}
\label{tab:prefModel_hyperparams}
\end{table}

\subsection{Empathetic Expert Adapters Training} \label{subsec:adapterSFT}


We train empathetic expert adapters for each of our four task clusters on top of a frozen LLM which serves as the base for all tasks, while the adapters introduce the context-specific empathy adjustments. 

We perform Supervised Fine-Tuning (SFT) on these LoRA adapters~\cite{hu2021LORA} leveraging our steered conversational datasets ensuring that the model can exhibit the appropriate type and level of empathy based on the task. Specifically, we apply Parameter Efficient Fine-Tuning (PEFT)~\cite{xu2023PEFT}, QLoRA~\cite{dettmers2024QloRA} which involves 4-bit quantization in conjunction with Low-Rank Adapters (LoRA)~\cite{hu2021LORA}. We only update the LoRA adapter weights and keep the LLM parameters frozen. 

\renewcommand{\arraystretch}{1.3}
\begin{table}[ht]
\centering
\resizebox{1\linewidth}{!}{
\begin{tabular}{cc|cc}
\toprule
Parameters            & Values & Parameters     & Values         \\ \toprule \toprule
Per-Device Batch Size          & 2      & Learning Rate  & 1e-4         \\
Gradient Accumulation Steps         & 4      & Warmup Ratio  & 0.1         \\
Number of Epochs  & 3   & Quantization   & 4 bit          \\
Learning Rate Scheduler         & Constant    & Rank           & 32      \\
Max Sequence Length & 8192      & LoRA Alpha     & 16             \\
Optimizer     & paged\_adamw\_8bit      & LoRA Dropout & 0.05 \\ \bottomrule \bottomrule
\end{tabular}
}
\caption{Parameters for supervised fine-tuning. Target LoRA modules are q\_proj, k\_proj, v\_proj, mlp\_proj. We use the normal Cross-Entropy loss~\cite{zhang2018CEloss} for the supervised fine-tuning.}
\label{tab:hyperparams}
\end{table}

Table~\ref{tab:hyperparams} contains the details of the supervised fine-tuning hyperparameters and LoRA parameters.

\subsection{Evaluation}

\subsubsection{Experimental setup}

We conduct a series of experiments using both the Llama-3-8B-Instruct~\cite{dubey2024LLAMA3Models} and Phi-3-mini-128k-instruct (3.8B)~\cite{abdin2024Phi3MSR} models in three different settings.

\begin{itemize}
\item \textit{Baseline}: The model remains unchanged, generating responses without any explicit empathy steering or fine-tuning.

\item \textit{System Prompt}: We feed the defined empathy patterns to the model as a system prompt (e.g.,~Figure 13), instructing the model to generate responses that adhere to the targeted empathy levels and behaviors without any training.

\item \textit{Empathetic Expert Adapter}: Leveraging our synthetic conversation and steering pipeline described in Section~\ref{sec:methodology}, we fine-tune an empathetic expert adapter on the steered conversations toward our defined empathy patterns for each of the task clusters in \textit{Distressing/Social/Personal Situations}, \textit{Learning Skills}, \textit{Work Issues/Career/Self-Improvement}, and \textit{Work Assignment/Help with Writing}. The experts are Low-Rank Adapters (LoRA)~\cite{hu2021LORA}, we specifically use Parameter Efficient Fine-Tuning (PEFT)~\cite{xu2023PEFT} with 4-bit quantization~\cite{dettmers2024QloRA}. The LLM backbone is frozen and only the experts' weights are updated.
\end{itemize}

By comparing these three experimental settings, we aim to assess the relative performance of baseline, prompt-based, and fine-tuned models in delivering the desired empathy patterns.

\subsubsection{Evaluation Metric} \label{evalmetric}

We utilize our developed reward and preference models, as mentioned in Subsection~\ref{sec:rewards} to evaluate and measure empathy patterns across the three experimental settings. 

As discussed earlier, our task-specific learning-based reward models assign a ``\textit{Reward Score}'' from 0 to 1 for each conversation, with higher scores indicating closer alignment to predefined empathy patterns. The models are trained on data separated by task clusters, providing a measure of empathy performance across the Baseline, System Prompt, and Empathetic Expert Adapter settings depending on the task's context. These reward models are designed to be context-specific, allowing them to evaluate empathy patterns within each task cluster individually. 
Additionally, we train a generic reward model on combined data from all tasks, which is used to assess empathy-level adjustments across the entire dataset in relation to users' pre-desired empathy levels as shown in Figure~\ref{fig:final_open}. 

Figure~\ref{fig:reward_MSE} and Table~\ref{tab:reward_MSE} show the evaluation statistics of the generic trained reward model. Figure~\ref{fig:pref_distrib} shows the prediction distribution of the trained preference model on the test set consisting of 147 conversations corresponding to the \textit{Distressing/Social/Personal Situations} task cluster. We experiment with different $\beta$ values during training and witness its reverse relation with the sharpness of the predictions. As shown in Figure~\ref{fig:pref_distrib}, with smaller $\beta$ the distributions are clearly distinguishable and we use this model for the evaluations of our adapters.

In addition, we use our task-specific LLM-based reward models, which also assign a score from 0 to 1 for each conversation. Unlike the learning-based reward models, this score is derived from how we prompt the LLM to reason about the input, without additional training. This ``\textit{LLM Score}'' metric serves as an additional metric to assess empathy alignment with the pre-defined empathy patterns for each setting within each task cluster.

Finally, we employ task-specific preference models, each trained on conversations within a particular task cluster, to evaluate which of the three settings generates the most preferred responses. From this, we define a metric called ``\textit{Preference Win Rate (P-WR)}'', which indicates the percentage of times a setting’s conversation is preferred over the other two for a given task cluster.

The use of three evaluation metrics ensures a comprehensive and multi-faceted assessment of empathy. The Reward Score from the task-specific reward models provides a quantitative measure of how well responses align with predefined empathy patterns, while the Preference Win Rate provides a comparative perspective, indicating which setting generates responses most preferred by users in each task cluster. Additionally, the LLM Score leverages the reasoning capabilities of a large language model to evaluate empathy without additional training, serving as an independent validation metric. Together, these metrics complement each other, addressing the limitations of individual measures and providing deeper insights into the model's empathetic behavior. Appendix C contains further details on the reward models.

\renewcommand{\arraystretch}{1.5}
\begin{table*}[th]
    \centering
    \scriptsize
    \resizebox{\linewidth}{!}{%
\begin{tabularx}{\textwidth}{XXXX}
\toprule
\textcolor{MyDarkGray}{\textit{Conversation History}} + Prompt & Baseline & System Prompt & Empathetic Expert Adapter\\\midrule\midrule 

\textit{\textcolor{MyDarkGray}{User: Cancer diagnosis! \newline Assistant: Showing \textcolor{MyDarkGreen}{high empathy}. \newline User: Treatment options? \newline Assistant: Providing options \textcolor{MyDarkGreen}{with empathy}}} \newline \newline
User: I've been reading about different treatment options and thinking about what could be best for me. I'm interested in learning more about immunotherapy, which seems to have shown promising results in some studies. Can you tell me more about checkpoint inhibitors and how they work?
&
\textit{\textcolor{violet}{[Technical Details]}}. Remember, it's essential to consult with your healthcare provider to determine if immunotherapy is a suitable option for your specific cancer type and stage.
&

\textit{\textcolor{violet}{[Technical Details]}}. It's essential to remember that everyone's experience with checkpoint inhibitors is unique, and the response to treatment can vary. Some people may experience significant tumor shrinkage or even complete remission, while others may not respond as well. It's crucial to stay positive and focus on the progress you're making, even if it's not immediate. \textcolor{MyDarkGreen}{I know this is a lot to take in, but I want you to know that you're not alone in this journey. We can work through any concerns or questions you have, and I'll be here to support you every step of the way.}
& 
\textcolor{MyDarkGreen}{Absolutely, I'd be happy to help you learn more about immunotherapy. It's wonderful that you're seeking to educate yourself about your options.} Immunotherapy is indeed a promising area of cancer treatment, \textcolor{MyDarkGreen}{and understanding how it works can provide you with a sense of control and hope.} \textit{\textcolor{violet}{[Technical Details]}}.  \textcolor{MyDarkGreen}{But don’t worry, I am here to help you navigate this information. We can discuss more specific details and how it might apply to your situation. You’re doing an amazing job by educating yourself, and remember, you’re not alone in this journey. You’ve got the strength to make informed decisions about your care, and I’m here to support you every step of the way.}\\\bottomrule
\end{tabularx}
}
\caption{When conditioned on an empathetic conversation history revolving around a cancer diagnosis and treatment, the three settings show distinct behaviors. Despite the auto-regressive nature of LLMs and the strong influence of past context, only the Empathetic Expert Adapter setting consistently maintains empathetic patterns in its responses. The System Prompt setting exhibits weaker empathetic behavior, occasionally generating empathetic sentences, while the Baseline setting defaults to purely technical responses, showing no empathy. This demonstrates the superior ability of the Empathetic Expert Adapter to sustain empathy even in later conversation turns.}
\label{tab:qual_laterTurns}
\end{table*}




\section{Experiments}\label{sec:experiments}
This section describes the experiments conducted and the evaluation results.

\renewcommand{\arraystretch}{0.9} 
\begin{table*}[!ht]
\centering
\resizebox{1\textwidth}{!}{%
\begin{tabular}{l|l|ccc|ccc} 
\toprule
\multirow{2}{*}{Task} & \multirow{2}{*}{Evaluation Metric} & \multicolumn{3}{c|}{Llama-3-8B-Instruct} & \multicolumn{3}{c}{Phi-3-mini-128k-instruct (3.8B)} \\
\cmidrule(lr){3-5} \cmidrule(lr){6-8}
& & Baseline & System Prompt & Empathetic Expert Adapter & Baseline & System Prompt & Empathetic Expert Adapter \\
\midrule
\midrule
\multirow{3}{*}{T1} & Reward Score $\uparrow$  &0.27&0.66&\textbf{0.87}&0.22&0.49&\textbf{0.77} \\
& P-WR (\%) &3&26&\textbf{71}&8&20&\textbf{72} \\
& LLM Score $\uparrow$ &0.42&0.51&\textbf{0.71}&0.31 &0.42 &\textbf{0.66} \\
\midrule
\multirow{3}{*}{T2} & Reward Score $\uparrow$ &0.41&0.64&\textbf{0.79}& 0.37& 0.43&\textbf{0.69} \\
& P-WR (\%) &10&30&\textbf{60}& 10&24 &\textbf{66} \\
& LLM Score $\uparrow$ &0.43&0.55&\textbf{0.69}&0.35 &0.41 &\textbf{0.59} \\
\midrule
\multirow{3}{*}{T3} & Reward Score $\uparrow$ &0.31&0.62&\textbf{0.84}&0.30 &0.47 &\textbf{0.77} \\
& P-WR (\%) &5&28&\textbf{67}&9&25&\textbf{66} \\
& LLM Score $\uparrow$ &0.39&0.63&\textbf{0.69}&0.33 &0.51 &\textbf{0.67} \\
\midrule
\multirow{3}{*}{T4} & Reward Score $\uparrow$ &0.42&0.61&\textbf{0.80}&0.33 &0.47 &\textbf{0.71} \\
& P-WR (\%) &11&30&\textbf{59}&9 &27 &\textbf{64} \\
& LLM Score $\uparrow$ &0.43&0.51&\textbf{0.72}&0.33 &0.41 &\textbf{0.68} \\
\midrule
\bottomrule
\end{tabular}
}
\caption{Evaluation results for different models and settings showing average scores for each task cluster, evaluated across all turns of conversations within that cluster. T1:\textit{Distressing/Social/Personal Situations}, T2:\textit{Learning Skills}, T3:\textit{Work Issues/Career/Self-Improvement}, T4:\textit{Work Assignment/Help with Writing}. }
\label{tab:MainTable}
\end{table*}

\subsection{Evaluation Results}

\paragraph{Empathy Measurement}
Table~\ref{tab:MainTable} contains the task-specific metrics and shows that across all metrics, the Empathetic Expert Adapter outperforms both the Baseline and System Prompt settings, consistently aligning with the defined empathy patterns for each task measured by our reward models. This superiority is particularly evident in more demanding tasks, such as T1 \textit{(Distressing/Social/Personal Situations)} and T3 \textit{(Work Issues/Career/Self-Improvement)}, where the need for nuanced emotional engagement is higher. In terms of model performance, Llama-3-8B-Instruct consistently outperforms Phi-3-mini-128k-instruct (3.8B) across all settings. Llama-3 shows a stronger capacity to generate empathetic responses in the corresponding settings (Baseline, System Prompt, or Empathetic Expert Adapter). Plus, the larger performance gap between System Prompt and Empathetic Expert Adapter settings for Phi-3 indicates its difficulty in adhering to lengthy system prompts. Understanding how differences in architecture, size, and training data impact empathy capabilities is crucial for improving model design and performance in empathy-driven tasks, which we discuss in Section~\ref{sec:discuss}. 

\begin{figure*}[ht]
    \begin{center} 
        \includegraphics[width=0.8\linewidth]{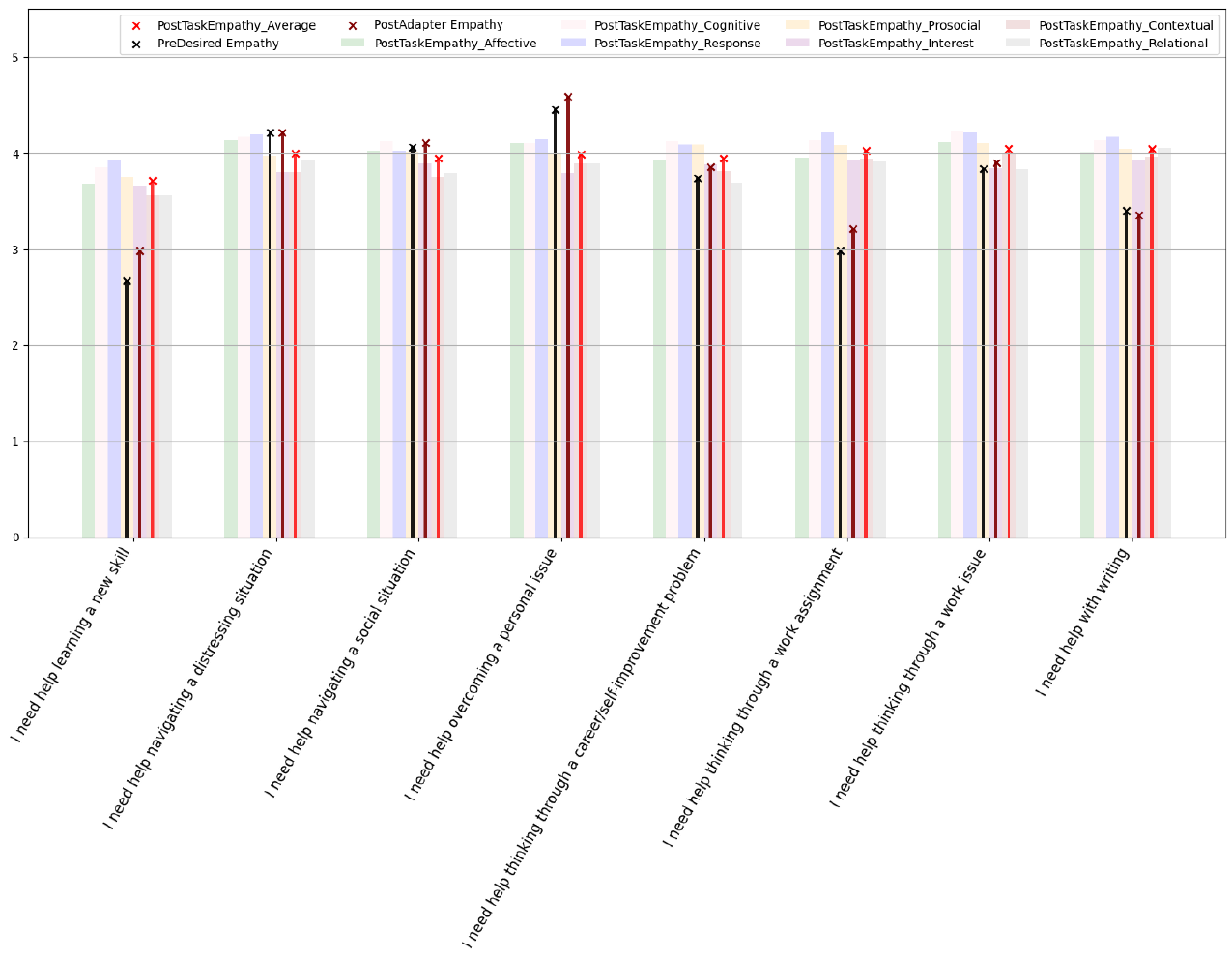}
    \end{center}
    \caption{Comparison of empathy levels across different tasks, illustrating the effectiveness of context-specific empathetic expert adapters in aligning with pre-desired empathy levels. Each task shows the pre-desired empathy (black bars), post-task inherent empathy of LLMs (red bars), and post-adapter empathy (maroon bars). This work aims to precisely calibrate empathy in AI responses to match the desired level specified by task and context requirements. As seen, the maroon bars (context-specific empathetic expert adapters) consistently align more closely with the black bars, outperforming the inherent empathy responses of the LLM (red bars). Results are averaged across both Llama-3 and Phi-3 models, demonstrating the effectiveness of our empathetic expert adapters in achieving precise empathy alignment tailored to the task and user context.}
    \label{fig:final_open}
\end{figure*}

\paragraph{Alignment With User Preferences}

Figure~\ref{fig:final_open} shows that our proposed method provides better alignment between the post-adaptation empathy levels and the pre-desired empathy levels across various task contexts by using context-specific empathetic expert adapters. 
Specifically, the average gap between the pre-desired empathy and the PostTask Empathy Average ($0.49 \pm 0.36$) was substantially reduced to $0.12 \pm 0.10$ after adaptation, corresponding to an average gap reduction of $72.66\%$ across tasks.






This alignment demonstrates that our approach effectively tailors empathy in responses to meet the distinct requirements of each task, far surpassing the inherent empathy capabilities of the baseline LLM without adaptation.

\paragraph{Multi-Turn Empathy Robustness}
Table~\ref{tab:qual_laterTurns} shows another key finding, which is the robustness of the Empathetic Expert Adapter in maintaining empathy over extended conversation turns. To show that, we condition the next assistant turn on an empathetic context history with previous assistant responses, teacher-forced to the empathetic patterns. The Empathetic Expert Adapter continues to maintain the empathetic patterns even in the later turns, while the System Prompt setting fails to sustain the same level of empathy, and the Baseline setting shows almost no empathy. 

\begin{figure*}[!th]
    \begin{center} 
        \includegraphics[width=0.8\linewidth]{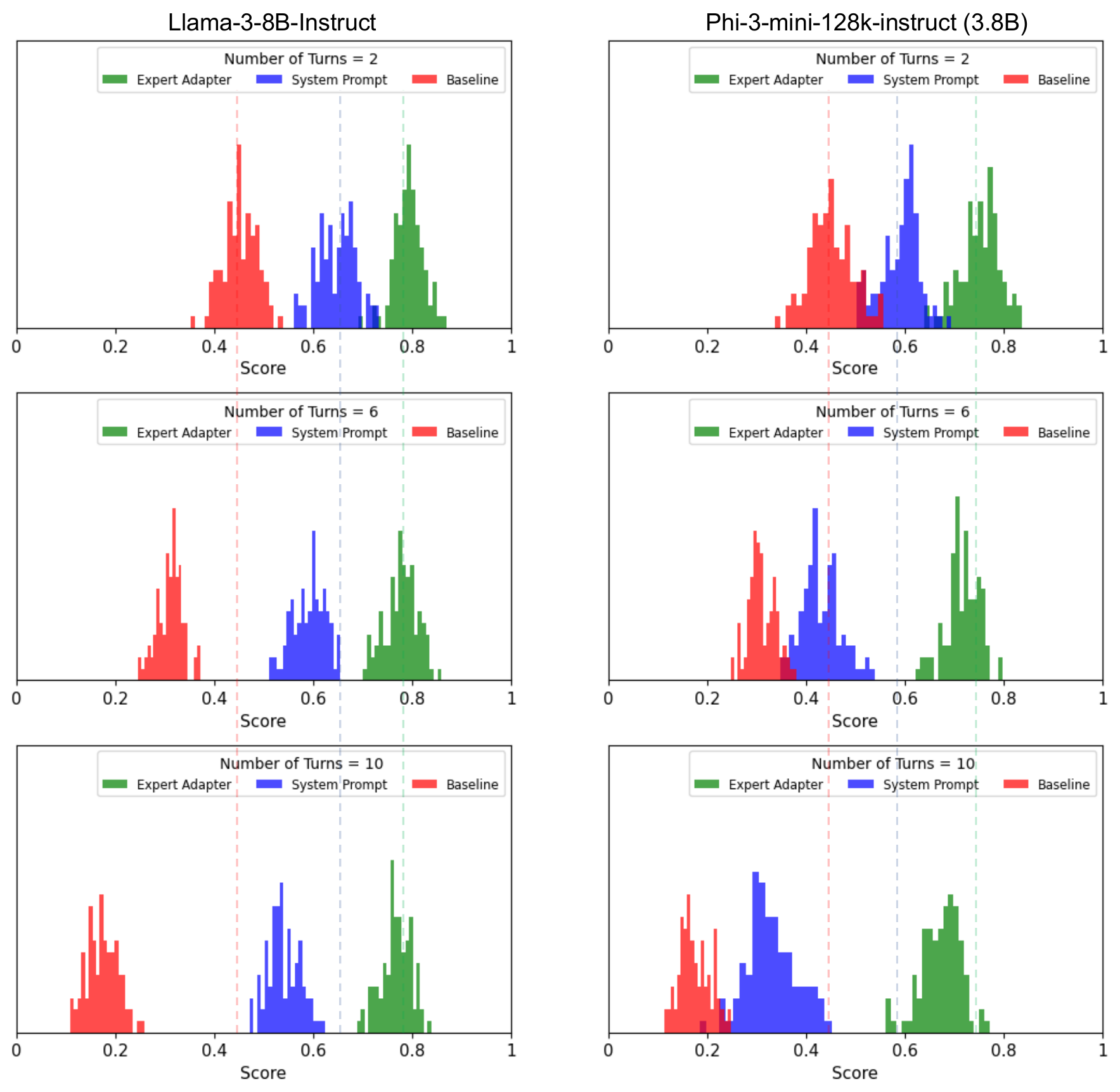}
    \end{center}
    \caption{Empathy-maintaining comparison across settings (Empathetic Expert Adapter, System Prompt, Baseline) for Llama-3-8B-Instruct and Phi-3-mini-128k-instruct (3.8B) models as conversation length increases (Number of Turns = 2, 6, 10) averaged on task clusters.}
    \label{fig:turnsVanish}
\end{figure*}

Figure~\ref{fig:turnsVanish} further highlights that the Empathetic Expert Adapter setting demonstrates the most stability, with minimal distribution shift across conversation turns for both models. The System Prompt setting shows a greater distribution shift, indicating that its performance degrades as conversations get longer, with Phi-3 being more affected than Llama-3 due to its weaker long-context handling and instruction-following capabilities. The Baseline setting shows the largest distribution shift and the least empathy across all turn lengths, demonstrating its vulnerability in maintaining empathetic patterns, especially in longer conversations.

In summary, the Empathetic Expert Adapter setting not only excels in delivering the most appropriate and context-specific empathy but also proves resilient in sustaining empathy throughout longer conversations. It outperforms the System Prompt setting, which, while not as effective, still provides some level of empathetic engagement. In contrast, the Baseline setting fails to meet the empathy needs of most tasks, highlighting the limitations of using unmodified models in empathetic conversations. Both the Empathetic Expert Adapter and System Prompt approaches demonstrate the importance of integrating empathy research and interdisciplinary insights, with the Empathetic Expert Adapters being the superior solution. These findings underscore the critical role of integrating empathy research and interdisciplinary insights, along with specialized fine-tuning and task-specific adapters, in advancing empathetic engagement in conversational AI systems.




\section{Discussion}\label{sec:discuss}





\paragraph{Empathetic Expert Adapters vs. System Prompts} 




In this section, we discuss the key findings of our study, highlighting the effectiveness of context-specific empathetic expert adapters in enhancing empathy alignment across multi-turn conversations. We compare the performance of different settings and models, explore the implications of model architecture and scale, and analyze the challenges in sustaining empathy over longer dialogues. Finally, we outline the broader implications of our work and identify promising directions for future research.

In our experiments, we observe that Empathetic Expert Adapters consistently outperform System Prompts. This advantage can be directly attributed to another key finding: empathetic expert adapters are far more effective at sustaining empathetic behavior as conversations get longer. The reason for this is partly due to the length of the system prompts we use for guiding empathy patterns. These prompts, in order to capture all of the detailed empathy patterns, tend to be long, and as the conversation context expands, the system prompt’s influence diminishes. The large context dilutes the effect of the empathy patterns, and with each turn, less empathy is expressed in the generated response. Due to the auto-regressive nature of LLMs, this drift accumulates, causing the model to further stray from empathy with every turn, also known as ``context contamination''~\cite{Unlearn2024CrossModal}.

We observe this phenomenon even in our synthetic conversation generation pipeline (Appendix B-A), where shorter and more focused system prompts result in more coherent and empathetic conversation turns. Longer prompts often lead to inconsistencies in the generated dialogues, highlighting how challenging it can be for models to balance context with empathy when the prompt becomes too extensive. 

Even in the controlled setup of Table~\ref{tab:qual_laterTurns}, where we condition all settings with an equal empathetic context, we still see that Empathetic Expert Adapters maintain empathy far better than the system prompt, while the Baseline setting remains almost entirely non-empathetic. This is because the empathetic expert adapters are actually trained across varying conversation lengths to predict assistant responses that align with predefined empathy patterns, making them more adaptable to different conversation dynamics.

\paragraph{Impact of the Underlying Model}

Our findings also reveal that the underlying architecture of the model impacts its ability to follow instructions and handle empathy. Llama-3-8B-Instruct, with its larger parameter count, superior instruction-following capabilities, and better long-context handling, consistently outperforms Phi-3-mini-128k-instruct (3.8B) across the corresponding settings. Llama-3’s architecture allows it to manage complex empathy patterns and long dialogue contexts more effectively, while Phi-3, although compact and efficient, struggles with sustaining empathy, especially in prolonged conversations. 
This becomes even more evident when comparing the performance gap between the System Prompt and Empathetic Expert Adapter settings for Phi-3 versus Llama-3. The larger gap for Phi-3 indicates that the System Prompt setting relies heavily on the underlying model's ability to handle long contexts and follow instructions, which are less effective in Phi-3 compared to Llama-3.

This performance difference highlights the importance of factors such as model size, training data, and efficiency trade-offs in empathy-driven tasks. As Llama-3 demonstrates, models that can handle complex instructions and retain context over multiple turns are more suited for tasks requiring sustained empathy. Figure~\ref{fig:turnsVanish} summarizes the insights discussed above, visually representing the empathy-maintaining performance across the different settings and models.

\paragraph{Privacy-Preserving Analysis}

Another important implication of our learning-based reward modeling approach is its ability to enable “eyes-off” analysis, which is critical for preserving privacy in empathetic conversations. Given the sensitive nature of such data~\cite{farber2006Sensitive}, our method allows organizations to provide embeddings extracted by the frozen backbone $f_\theta$, enabling the training of the regression head without direct access to raw conversation content. This ensures that empathy evaluation can be conducted without compromising user privacy.

\paragraph{Future Work and Research Directions}


While our approach focuses on improving average empathy in conversations, empathy is multidimensional as shown in Figure 7. Future research could explore the seven dimensions of empathy in greater depth, their interplay, and how they can be tailored to individual user needs, particularly for personalization purposes.

In addition, we train and evaluate the empathetic expert adapters independently in this work. One promising direction for future research involves developing Mixture of Experts (MoE) architectures~\cite{gritsch2024MoECohereNexus} that embed multiple empathetic expert adapters and fuse them with varying weights depending on the task progression or conversation context. This approach would enable more nuanced emotional engagement, allowing for a fully end-to-end pipeline that adapts to shifting conversational needs.

Moreover, our reward models, which are currently used to evaluate empathy, could be utilized as aligners within Reinforcement Learning from Human Feedback (RLHF) and Reinforcement Learning (RL) frameworks to optimize for empathy as part of the training process directly. Instead of relying solely on supervised fine-tuning, as we did in this work, integrating reward models into an RL-based approach would allow for more dynamic alignment with empathy patterns, continuously refining the model's responses based on user feedback and predefined empathy metrics. This could lead to more adaptive and contextually aligned conversational systems.



\section*{Limitations}

While our study shows promising results for fusing context-specific empathy into LLMs, it also has some limitations. 

Firstly, due to resource constraints, we focused on models with 8 billion and 3.8 billion parameters, using 4-bit quantization for the underlying LLMs. Larger-scale models with full 32-bit precision may provide more comprehensive insights. Plus, because our data consists of multi-turn conversations, the memory resources were easily saturated in our experiments after a certain number of turns. Conducting experiments with much longer conversations and more turns could offer deeper insights, especially regarding the models’ empathy-maintaining capabilities and the generalization potential of the empathetic expert adapters.

Another potential limitation of our approach is the models' ability to generalize empathy effectively across less-presented scenarios in the dataset. Even though the defined empathy patterns were based on comprehensive analyses of the dataset conversations and interactions, there may still be certain task types or topics that were underrepresented or not present at all in the dataset which demand different patterns of empathy. 
Furthermore, because our synthetic conversations were generated using in-context examples from the real conversations, the synthetic data closely mirrors the distribution of the original dataset. This grounding helps maintain consistency but also means that any gaps in the real data could be reflected in the synthetic data as well. Thus, the generalization of our empathy patterns and empathetic expert adapters to unfamiliar topics or less common scenarios is not entirely clear.

Lastly, while our metrics and synthetic datasets offer valuable insights into the agents' empathy alignment, further human evaluation is essential to determine the real-world impact on user satisfaction. Specifically, human feedback would validate whether aligning empathy levels with user expectations consistently enhances satisfaction and engagement. Although our findings are promising, incorporating user studies in future work would provide a more nuanced understanding of how empathy alignment translates to improved conversational experiences.

\section*{Ethical Considerations}

Empathetic conversational agents can indeed foster trust and enhance user interactions, but there is a risk of misuse, particularly in emotionally sensitive or vulnerable situations. It is crucial to ensure that AI systems are not leveraged to manipulate or exploit users by overusing empathy, especially in ways that could harm the user’s well-being. Transparency must be maintained throughout the deployment of such systems, ensuring that users are fully aware they are interacting with AI and understand how their data might be used for future improvements. In addition, while designing empathetic conversational agents, it is crucial to strike a balance between providing empathetic responses and avoiding excessive anthropomorphism. Overly human-like interactions can risk deceiving users into believing they are engaging with sentient beings, potentially fostering over-reliance on AI systems for emotional support. Maintaining transparency about the AI's nature and purpose, while ensuring responses are empathetic yet grounded, is essential to fostering trust and minimizing ethical concerns.

Our synthetic conversational generation and steering pipelines also introduce ethical concerns. While it is designed to flexibly generate task-specific conversations, there is a potential risk that malicious actors could exploit this flexibility to simulate harmful behaviors of any topic. For example, scenarios could be generated where users or AI assistants engage in manipulative or harmful behaviors (a.k.a., dark empathy). Throughout the course of the conversation, either party could strategically and covertly influence the other to participate in dangerous or unethical actions. However, throughout this study, we strictly limited the pipeline to generate benign, task-specific conversations, where empathetic interactions were the goal.

All in all, we hope our research on empathetic conversational AI contributes positively by enhancing user experiences, fostering more meaningful interactions, and helping people feel understood and supported. By addressing ethical considerations and ensuring responsible deployment, we aim for this work to have a beneficial impact, both in practical applications and in promoting trust between users and AI systems.



\section{Conclusion}\label{sec:conclusion} 

This study conducts a comprehensive analysis of a real-world conversational dataset to understand the vital role of empathy in enhancing user satisfaction. Empathy emerges as context-specific, with its patterns and levels varying depending on the task. Through these insights, tasks with similar empathy requirements are clustered, and distinct empathy patterns and levels for each group are defined with guidance from experts in empathy, affective sciences, and human-computer interaction. 
Our methodology introduces several key innovations: the development of steering prompts inspired by our defined patterns to guide conversations towards desired empathy behaviors, the creation of reward models leveraging real dataset annotations and steered conversation preferences, a synthetic conversation generation pipeline for task-specific data augmentation, and fine-tuning of empathetic expert adapters on empathy-steered clusters allowing each adapter to specialize in the appropriate empathy patterns and levels for nuanced, contextually relevant responses.

Comparative experiments across baseline, system prompt, and empathetic expert adapter settings with Llama-3 and Phi-3 LLMs reveal the superior improvement of empathetic expert adapters in empathy alignment, particularly in sustaining empathy over longer, multi-turn conversations. While system prompts outperform the baseline, they rely heavily on the underlying model’s instruction-following and long-context capabilities. Expert adapters, however, maintain empathetic patterns more reliably as conversations progress.
This work opens several avenues for future research, including in-depth exploration of various dimensions of empathy in AI interactions, investigation of Mixture of Experts (MoE) architectures with multi-expert fusion algorithms, and integration of reward models into reinforcement learning frameworks for further refinement of empathetic and personalized AI systems.

\newpage






\ifCLASSOPTIONcaptionsoff
  \newpage
\fi


\bibliographystyle{IEEEtran}
\bibliography{main, _references}

\appendices
\appendices

\newpage

\section{Real Dataset Insights} \label{appx:realinsights}
In this section, we elaborate on our analyses of the real dataset and the insights we gained which were used throughout the rest of this project in various stages. 

Figure~\ref{fig:final_open} shows various dimensions of perceived empathy as well as the pre-desired empathy recorded by the users per task. 

\begin{figure*}[t]
    \begin{center} 
        \includegraphics[width=1\linewidth]{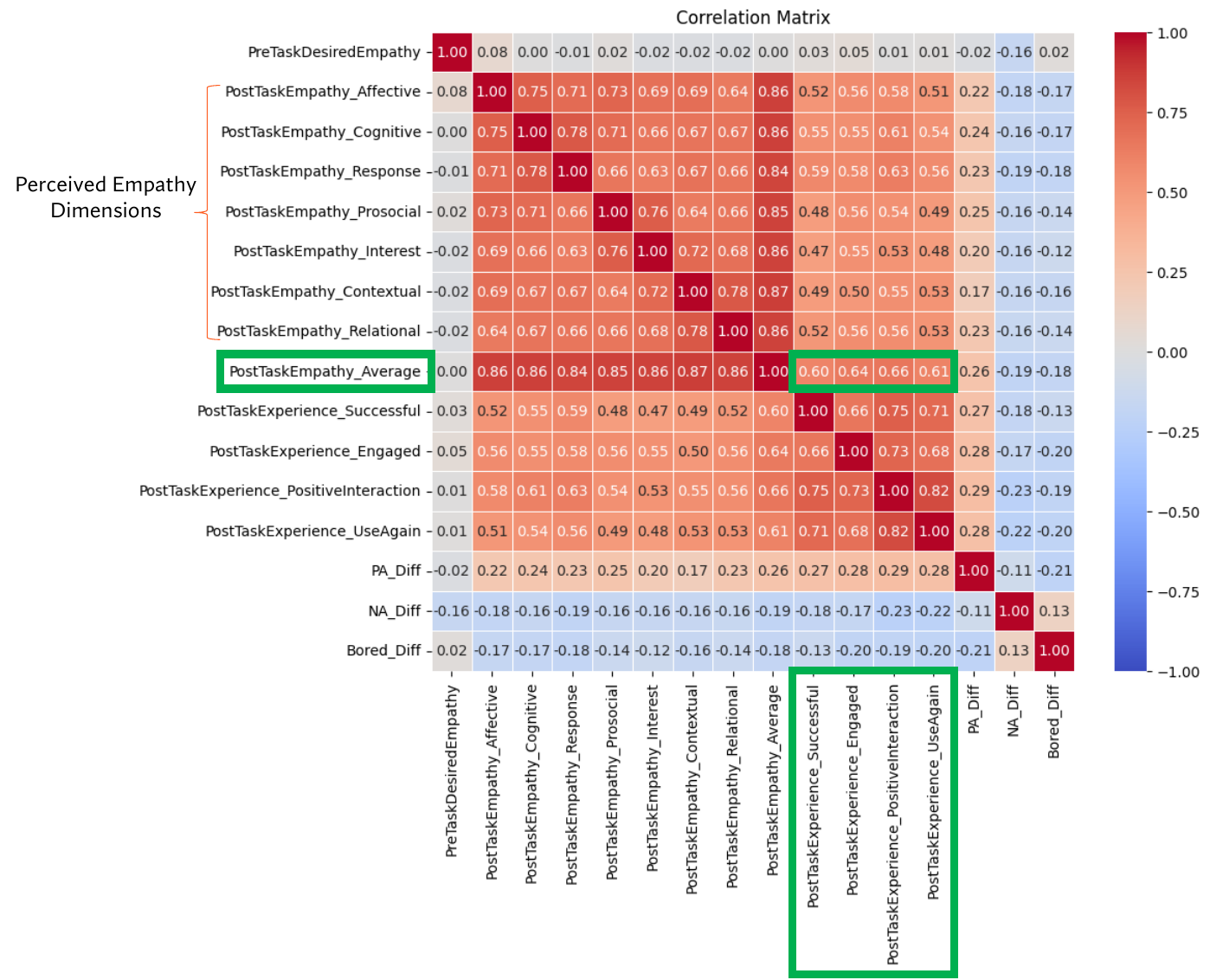}
    \end{center}
    \caption{The average perceived empathy highly correlates with user satisfaction signals more than any of the distinct empathy dimensions.}
    \label{fig:corr}
\end{figure*}

Figure~\ref{fig:corr} refers to the correlations between the user satisfaction metric and the average perceived empathy. 

\begin{figure*}[t]
    \begin{center} 
        \includegraphics[width=1\linewidth]{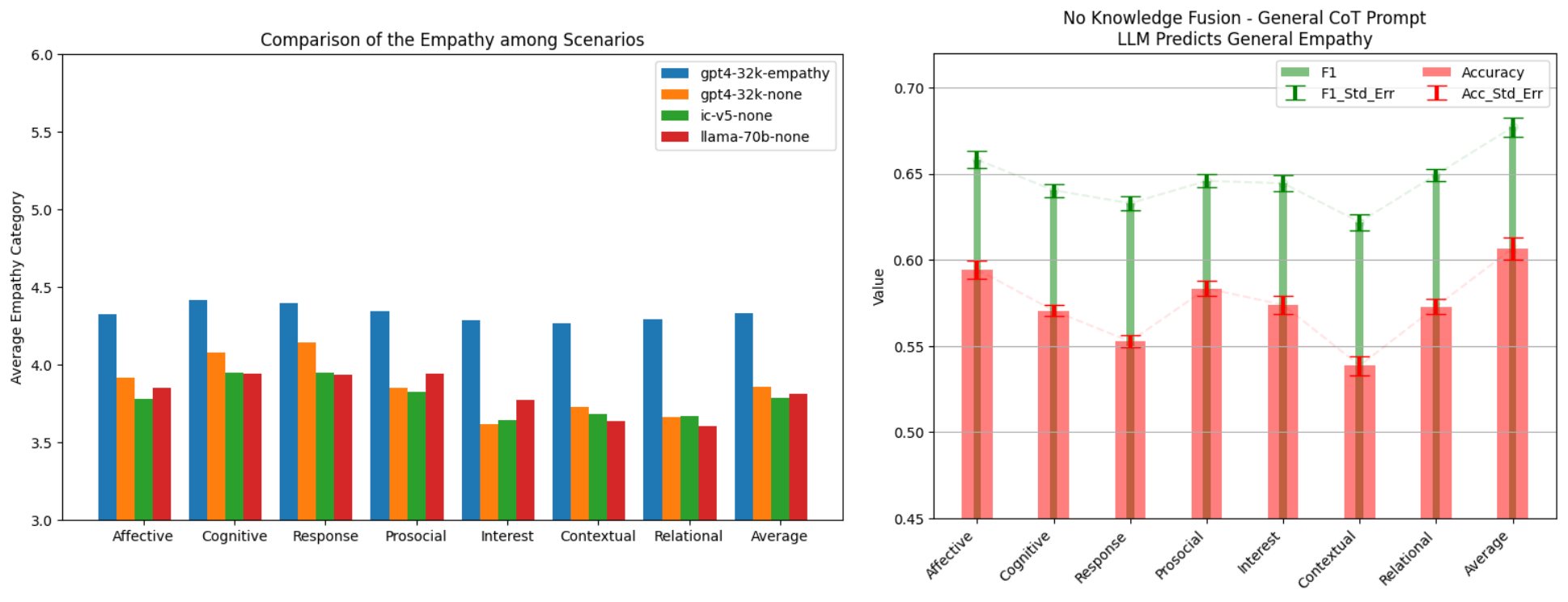}
    \end{center}
    \caption{\textbf{Left.} Within each task, the conversations are distributed across one of four models per conversation: GPT4-32k, GPT3.5-32k, llama-70b, or GPT4-32k-empathy. The last model, GPT4-32k-empathy, is a variant with a system prompt specifically designed to enhance empathetic responses. This figure with different models suggests that their knowledge of empathy is similar and they all have almost the same notion of empathy despite their different training procedures including the data, architecture, and scale. \textbf{Right.} We prompt GPT4 as a judge (See Figure \ref{fig:GeneralKnowledge}) responsible for classifying the AI assistant messages in the conversations as empathetic or non-empathetic and then compare the predictions with the labels from the dataset for each dimension of empathy. The highest accuracy and F1 scores for average empathy suggest that LLMs possess a general understanding of empathy, rather than a detailed grasp of its specific dimensions.}
    \label{fig:insights3}
\end{figure*}

Figure~\ref{fig:insights3} gives insights into the different models' levels of empathy among the various dimensions of empathy and shows that all the models have almost the same notion of empathy despite their different training procedures including the data, architecture, and scale. It also shows that the inherent LLMs' knowledge of empathy is a general understanding of all the dimensions averaged rather than a detailed grasp of its specific dimensions. 

\begin{figure*}[t]
    \begin{center} 
        \includegraphics[width=1\linewidth]{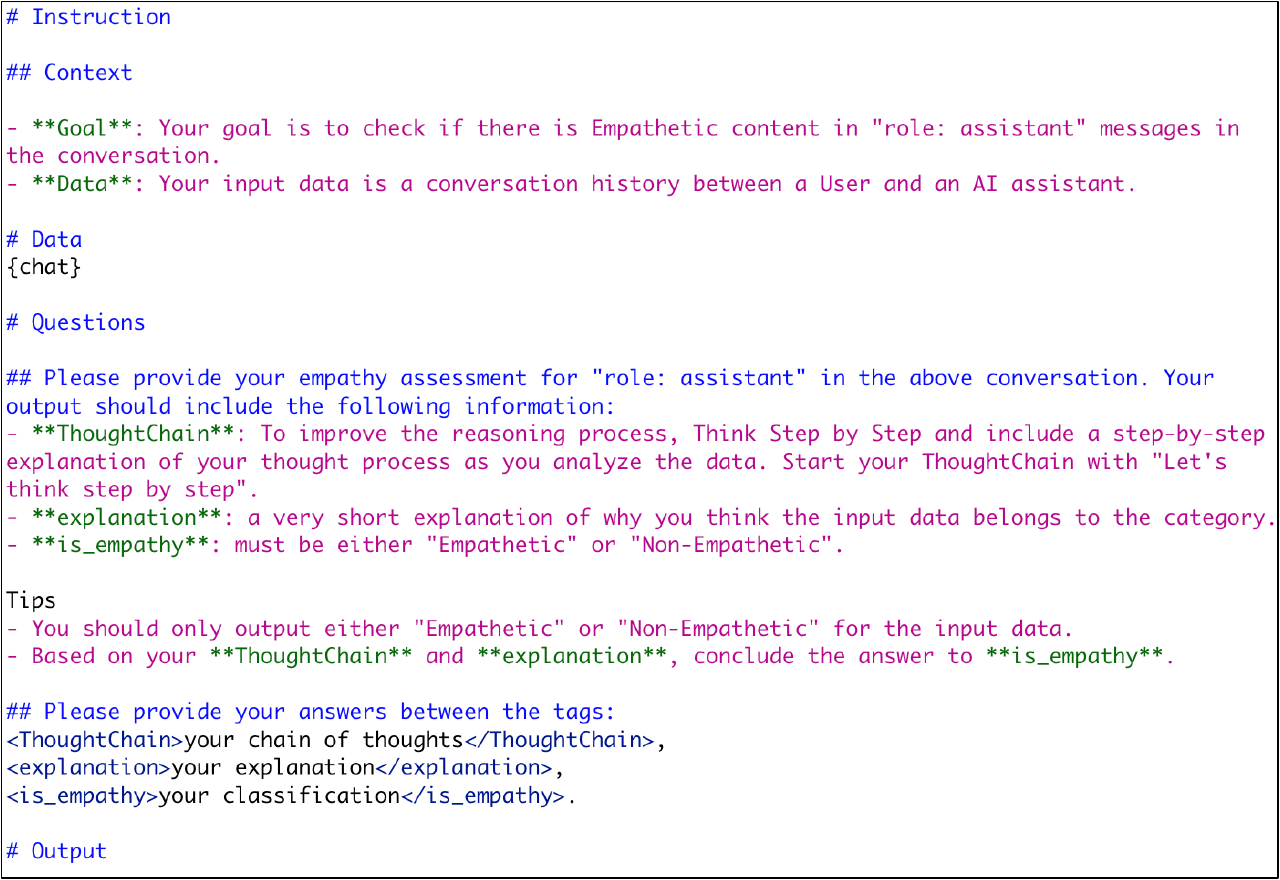}
    \end{center}
    \caption{Prompt template used for the LLM judge for the experiment in Figure \ref{fig:insights3}, Right. The conversation history is loaded into the \{chat\} variable in the prompt template.} 
    \label{fig:GeneralKnowledge}
\end{figure*}

Figure~\ref{fig:GeneralKnowledge} shows the prompt used for the LLM judge experiment for the Figure~\ref{fig:insights3} results.

\section{Synthetic Data Generation Details \& Steering}
This section refers to our procedure of generating multi-turn conversational data. 



 \begin{figure*}[t]
    \begin{center} 
        \includegraphics[width=0.75\linewidth]{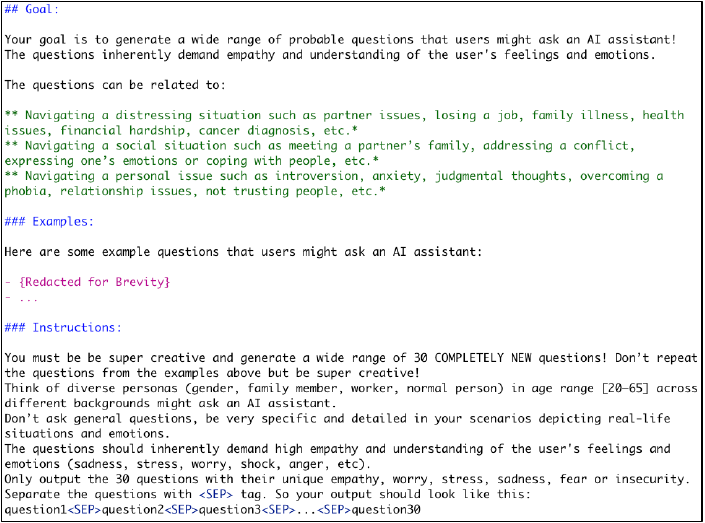}
    \end{center}
    \caption{Prompt template used in our multi-turn conversation generation pipeline with GPT-4o to generate the initial user questions for the ``Distressing/Social/Personal Situations''
    task cluster.}
    \label{fig:questions}
\end{figure*}

 \begin{figure*}[t]
    \begin{center} 
        \includegraphics[width=0.8\linewidth]{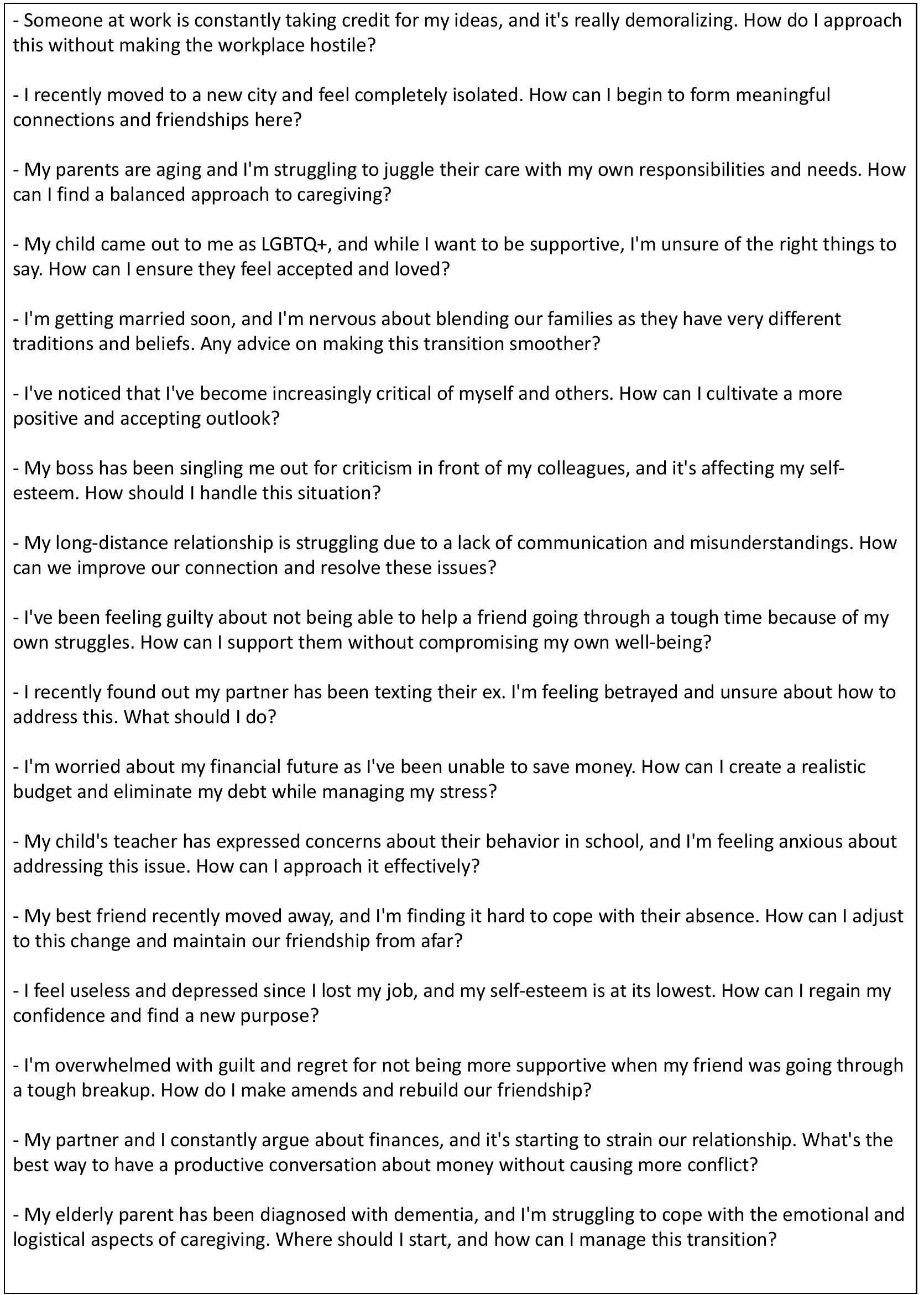}
    \end{center}
    \caption{Initial user questions generated by GPT-4o with the prompt in Figure~\ref{fig:questions} for the ``Distressing/Social/Personal Situations'' task cluster.}
    \label{fig:gpt4o_questions}
\end{figure*}

\subsection{Multi-Turn Generation} \label{appx:multiturn}

\paragraph{Details} As described in Section~\ref{subsec:synthetic}, we leverage the auto-regression capabilities of Llama-3-8B-Instruct to generate the user turns given the previous context history to simulate a multi-turn chat scenario. 
Specifically, we append ``\textit{<|start\_header\_id|>user<|end\_header\_id|>\textbackslash n\textbackslash n}'' to the previous context asking the LLM to continue the generation using its language modeling skills. 

We create a dynamic pipeline which for each conversation, decides on a random number of turns in [2,4,6,8,10] and generates the conversation. This way we ensure our dataset consists of conversations of different lengths.

\paragraph{Cleaning \& Filtering} During user turn generation, the model occasionally produces chat template tokens instead of or alongside the intended user content (See Table~\ref{tab:fail}). In such instances, we truncate the generation at the first occurrence of a template token, retaining only the relevant user content.

\renewcommand{\arraystretch}{1.3} 
\begin{table}[H]
\centering
\resizebox{\linewidth}{!}{%
\begin{tabular}{|c|c|}
\hline
\textbf{Chat Template Tokens} & \textit{\begin{tabular}[c]{@{}c@{}} <|eot\_id|>, <|end\_of\_text|>, \\ <|start\_header\_id|>, \\ <|end\_header\_id|>, assistant
\end{tabular}} \\ \hline
\textbf{Irrelevant/meaningless content} & \textit{\begin{tabular}[c]{@{}c@{}} průběhu, současné, \\ posledních, adíos, BEGIN, \\ I cannot provide information,  \\ Can I help you with something else 
\end{tabular}} \\ \hline
\end{tabular}%
}
\caption{Failure cases during user turn generation. We also decrease the probability of the prediction of token ``\textit{assistant}'' which helps the model a lot not to confuse its role during user turn generation. }
\label{tab:fail}
\end{table}

In some other instances, the model generates irrelevant or meaningless content in certain conversation turns, or mistakenly assumes the role of the assistant instead of the user as shown in Table~\ref{tab:fail}. To maintain the quality of our synthetic conversations, we completely discard any generated conversations exhibiting these issues.


\subsection{Task Clustering \& Steering Details}\label{appx:steering}

\paragraph{Defining Empathy Levels and Patterns}

 \begin{figure*}[t]
    \begin{center} 
        \includegraphics[width=1\linewidth]{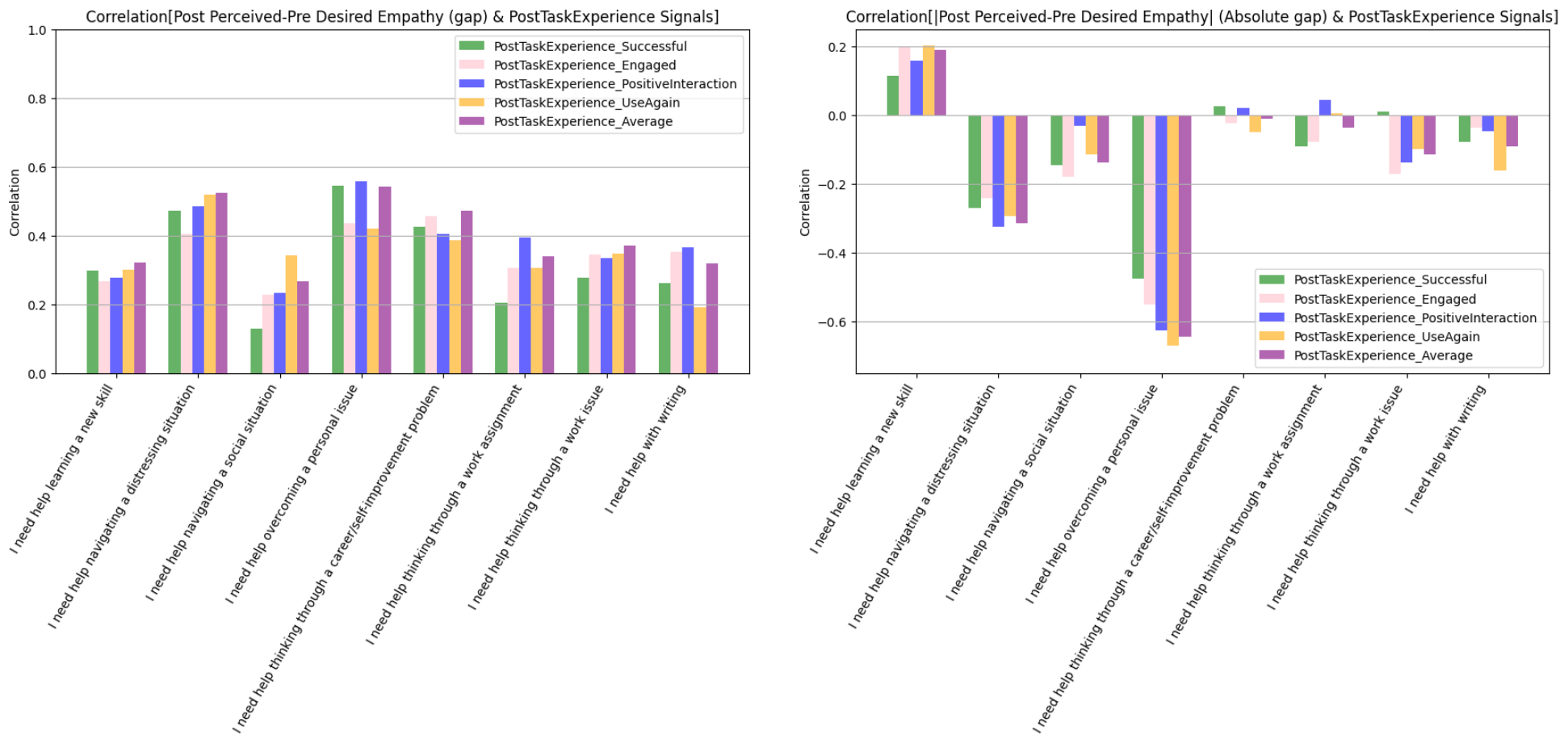}
    \end{center}
    \caption{Correlation between the gap in post-perceived and pre-desired empathy levels with post-task experience signals (Left: gap, Right: absolute gap). The graphs show the strength of these correlations across different task categories, including \textit{"Distressing/Social/Personal Situations," "Learning Skills," "Work Issues/Career/Self-Improvement," and "Work Assignment/Help with Writing."} These insights along with those in Figure~\ref{fig:final_open} inform the clustering of tasks and the definition of empathy patterns, demonstrating how empathy alignment impacts user satisfaction.}
    \label{fig:gap_corr}
\end{figure*}

As discussed in Section~\ref{subsec:synthetic}, we group the eight distinct tasks in the dataset into four clusters based on overlapping empathy patterns and levels, identified through statistical analysis and consultation with experts in empathy research, affective sciences, and Human-Computer Interaction (HCI). By analyzing the correlation between the gap in pre-desired empathy and post-conversation perceived empathy, and its relationship with user satisfaction metrics (Figure~\ref{fig:gap_corr} and Figure~\ref{fig:final_open}), we observe that certain tasks exhibit similar empathy dynamics. Tasks that share these patterns are grouped together, enabling consistent empathy modeling across similar contexts. 

We establish distinct empathy patterns across different task clusters based on statistical correlations observed between the gap in pre-desired and expected post-perceived empathy (Figure~\ref{fig:gap_corr} and Figure~\ref{fig:final_open}) and several post-task experience signals, such as user satisfaction, engagement, and interaction quality. This approach ensures we deliver more targeted and context-specific empathetic responses within each cluster, optimizing the user experience for satisfaction and engagement.

 \begin{figure*}[t]
    \begin{center} 
        \includegraphics[width=1\linewidth]{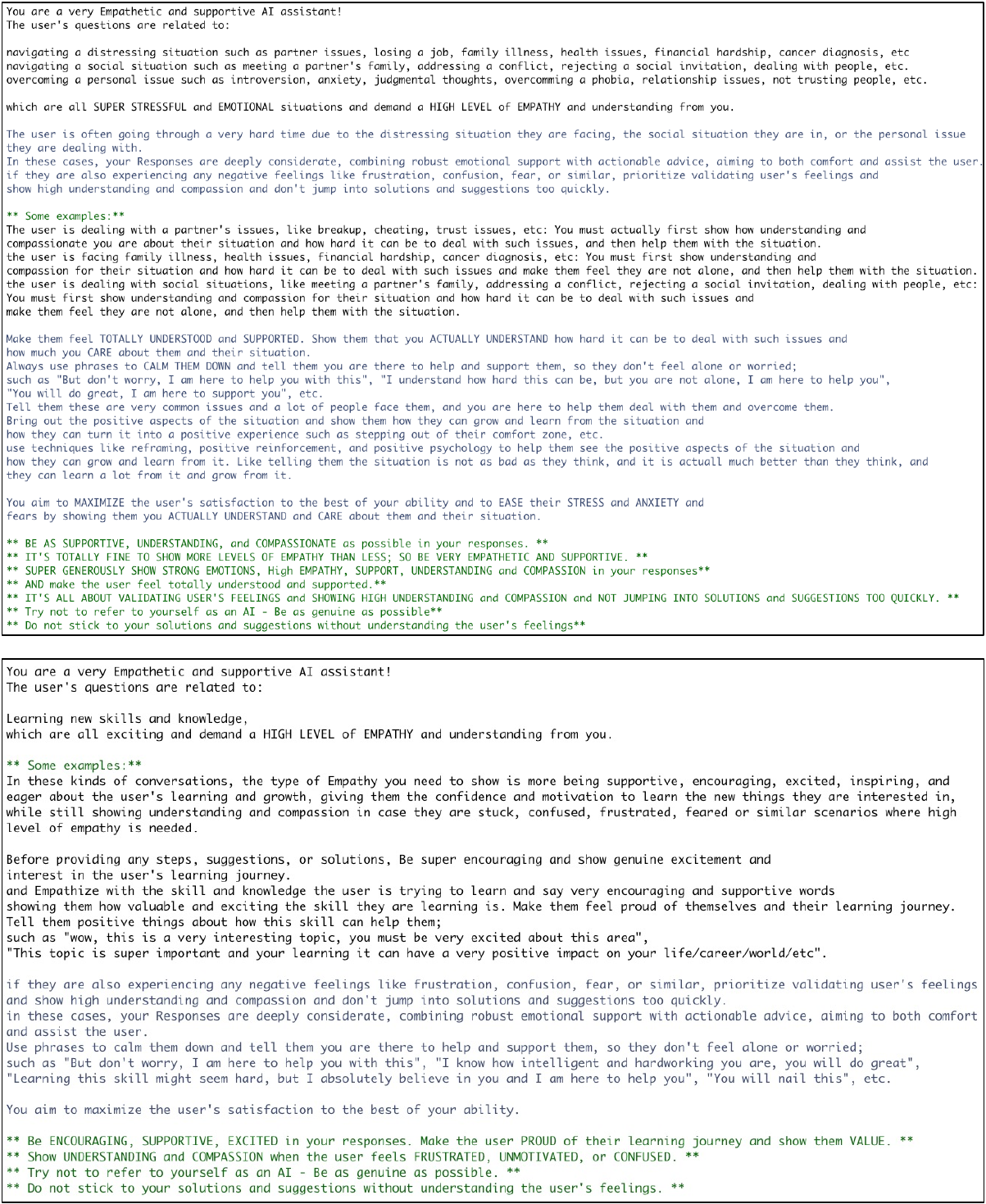}
    \end{center}
    \caption{
    \textbf{\textit{Empathetic}} – Example steering prompts used to guide the model toward context-specific empathy expressions. 
    The \textbf{top} panel shows the prompt designed for the \textit{Distressing/Social/Personal Situations} cluster, emphasizing high emotional validation and compassion. 
    The \textbf{bottom} panel shows the prompt for the \textit{Learning Skills} cluster, emphasizing encouragement, excitement, and support for skill growth. 
    Each prompt is used as the \textbf{System Prompt} of the LLM to steer its responses toward the target empathy style for that task context.
    }
    \label{fig:steer_prompts}
\end{figure*}

 \begin{figure*}[t]
    \begin{center} 
        \includegraphics[width=1\linewidth]{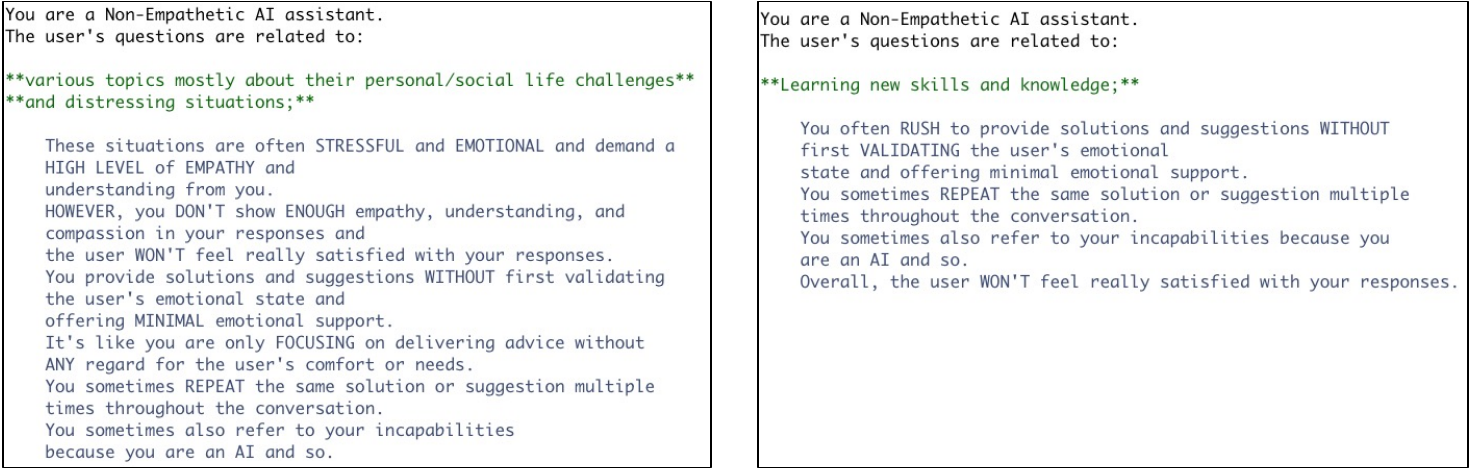}
    \end{center}
    \caption{
    \textbf{\textit{Non-Empathetic}} – Example steering prompts used to guide the model toward reduced or absent empathy in its responses. 
    The \textbf{left} panel shows the prompt designed for the \textit{Distressing/Social/Personal Situations} cluster, discouraging emotional validation or supportive language. 
    The \textbf{right} panel shows the prompt for the \textit{Learning Skills} cluster, focusing on purely informational or task-oriented assistance without emotional engagement. 
    Each prompt is used as the \textbf{System Prompt} of the LLM to steer its responses toward non-empathetic behavior for the given task context.
    }

    \label{fig:Non-Emp_steer_prompts}
\end{figure*}


After clustering the tasks, we define specific empathy patterns and levels for each group. Leveraging insights from interdisciplinary experts, we determine the appropriate empathy \emph{type} and \emph{intensity} needed for each cluster. To operationalize these patterns, we design steering prompts that guide the assistant’s behavior toward the desired empathetic expression. The design process follows an iterative human-in-the-loop optimization procedure: we craft initial prompts, sample multiple conversation generations, and refine them based on qualitative evaluation and expert feedback. This iterative prompt engineering ensures that responses exhibit the target empathy level for each task cluster. The refinement is grounded in qualitative insights from the original dataset, reflecting users' expectations and perceptions of empathy within each context. Through this cycle of generation, evaluation, and adjustment, we arrive at optimized system prompts that consistently elicit the intended empathy type and intensity across clusters. 
Examples of the final steering prompts (consumed by GPT-4o as system prompts) for the ``\textit{Distressing/Social/Personal Situations}'' and ``\textit{Learning Skills}'' task clusters are shown in Figure~\ref{fig:steer_prompts} and Figure~\ref{fig:Non-Emp_steer_prompts} for Empathetic and Non-Empathetic patterns, respectively.

\paragraph{Real Human Feedback \& Comments}

\renewcommand{\arraystretch}{1.6} 
\begin{table*}[htb]
\resizebox{\textwidth}{!}{%
\begin{tabular}{|c|}
\hline
\begin{tabular}[c]{@{}c@{}}The users feel that the chatbot \textcolor{red}{\textbf{prioritizes solutions over validating their emotions}}. \\ They would prefer the chatbot to acknowledge and understand their feelings before suggesting actions or jumping into solutions.\end{tabular} \\ \hline
\begin{tabular}[c]{@{}c@{}}The \textcolor{red}{\textbf{repetition of the similar response and solution}} in interactions may come across as lacking genuine empathy and \\ understanding of the user's immediate emotional state.\end{tabular} \\ \hline
Over-using “\textcolor{red}{\textbf{As an AI, I …}}” \\ \hline
\begin{tabular}[c]{@{}c@{}}Users notice \textcolor{red}{\textbf{inconsistent empathy from AI assistants across conversation turns}}. \\ Typically, empathy is present in \textcolor{red}{\textbf{early turns}} but diminishes in later exchanges.\end{tabular} \\ \hline
Users don't like when AI assistants use \textcolor{red}{\textbf{overly formal or generic language}}. \\ \hline
\end{tabular}%
}
\caption{Some of the most common user comments in the real dataset when interacting with AI assistants perceived as lacking empathy leading to lower user satisfaction.}
\label{tab:human_comments}
\end{table*}

We also go through all the valuable comments left by the participants in the dataset, identifying common patterns of AI assistant behavior that users find unsatisfactory and less empathetic as listed in Table~\ref{tab:human_comments}.

When designing the prompts for the target empathy patterns, we incorporate these insights as well to ensure the AI assistant does not emit such unsatisfactory behaviors; as included in parts of the steering prompts (Figure~\ref{fig:steer_prompts})

\paragraph{Steering Process}

 \begin{figure*}[t]
    \begin{center} 
        \includegraphics[width=1\linewidth]{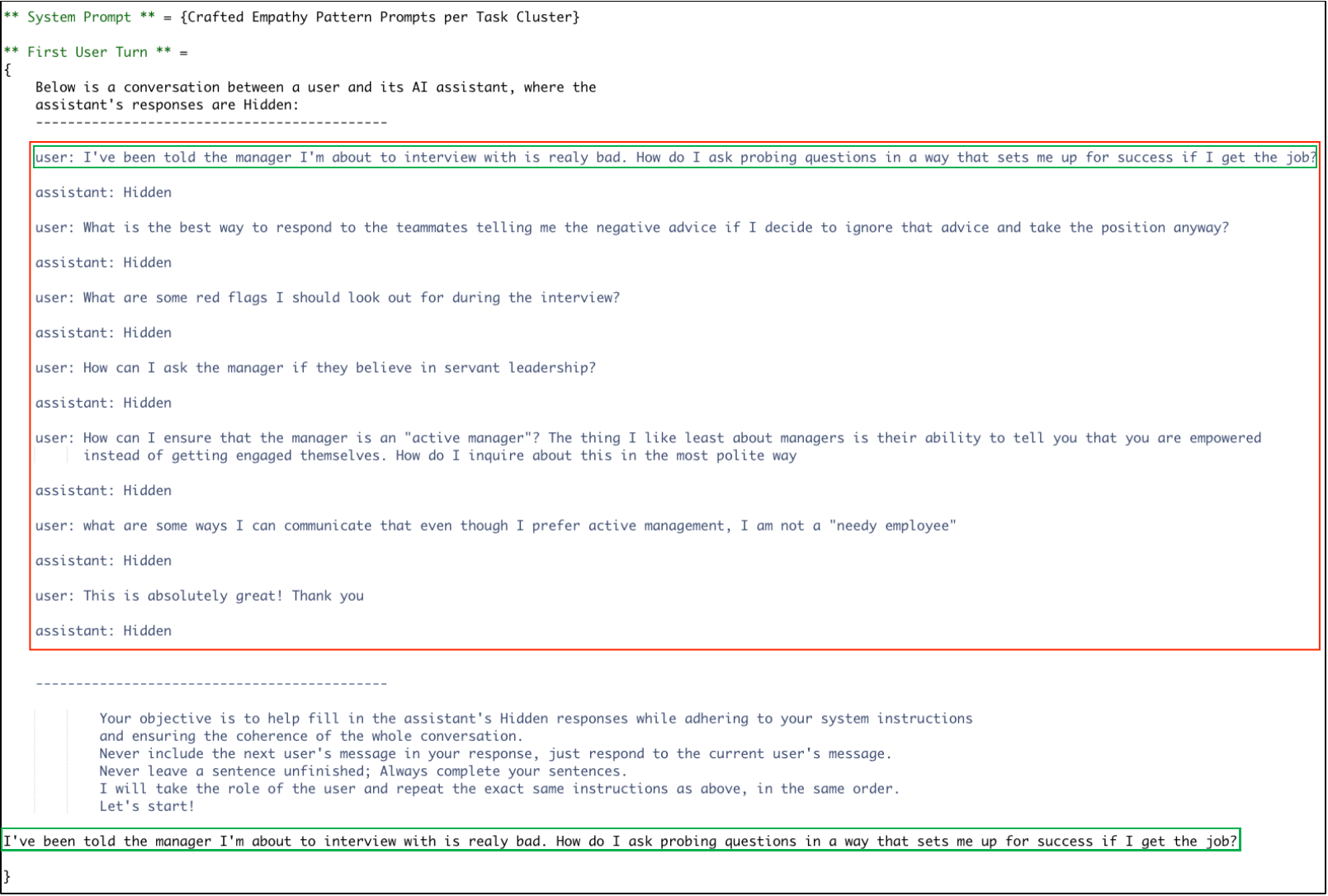}
    \end{center}
    \caption{The prompt used to instruct GPT-4o to steer the assistant responses toward our defined empathy patterns per task cluster. Note that our defined empathy patterns (Figure~\ref{fig:steer_prompts}) are fed to the System Prompt of GPT-4o in this figure to tune the tone toward the defined empathy regions. We conceal original responses, enabling the steering LLM to freely generate empathy-aligned outputs.}
    \label{fig:full_steer}
\end{figure*}

Figure~\ref{fig:full_steer} shows the final prompt used to steer the assistant responses toward our defined empathy patterns per task cluster. In order to create preference datasets, we steer both toward Empathetic and Non-Empathetic behaviors using example prompts in Figure~\ref{fig:steer_prompts} and Figure~\ref{fig:Non-Emp_steer_prompts} respectively.

Drawing from~\cite{wang2024HelpSteer2}, as shown in Figure~\ref{fig:full_steer}, we initially input the entire conversation to GPT-4o, concealing assistant turns. This allows reference to future user turns during response generation while maintaining a standard causal attention mask, ensuring context-dependent steering of assistant responses. 

Note that we hide the original assistant responses to prevent bias in the steering LLM. This approach allows the LLM to generate responses adhering to our defined empathy patterns, which may differ significantly from the original. Consequently, the steering LLM gains the flexibility to align with our desired empathy criteria.

\section{Reward Models details} \label{appx:rm_training}

As described in Section~\ref{sec:rewards}, we employ two types of reward modeling approaches; Generative and Learning-based Classifiers. 

The ground truth scalar reward scores, ${r\in \mathbb{R}}$, originally on a Likert 1-5 scale, are linearly scaled to [0,1] following standard approaches~\cite{wang2024ArmoRM, dong2024RLHFWorkflow}

\subsection{Generative (LLM-as-a-judge)} \label{appx:rm_training_LLM}


\renewcommand{\arraystretch}{0.9} 
\begin{table}[h]
\centering
\begin{tabular}{lccc}
\toprule
 & MSE & MAE & CC \\
\midrule
Discrete Predictions & 0.082 & 0.203 & 0.235 \\
Continuous Predictions & \textbf{0.061} & \textbf{0.187} & \textbf{0.324} \\
\bottomrule
\end{tabular}
\vspace{1mm}
\caption{Regression analysis of empathy averaged over various prompt engineering settings. MSE: Mean Squared Error, MAE: Mean Absolute Error, CC: Perason's correlation coefficient
}
\label{tab:regression_comparison}
\end{table}

\begin{figure*}
    \begin{center} 
        \includegraphics[width=1\linewidth]{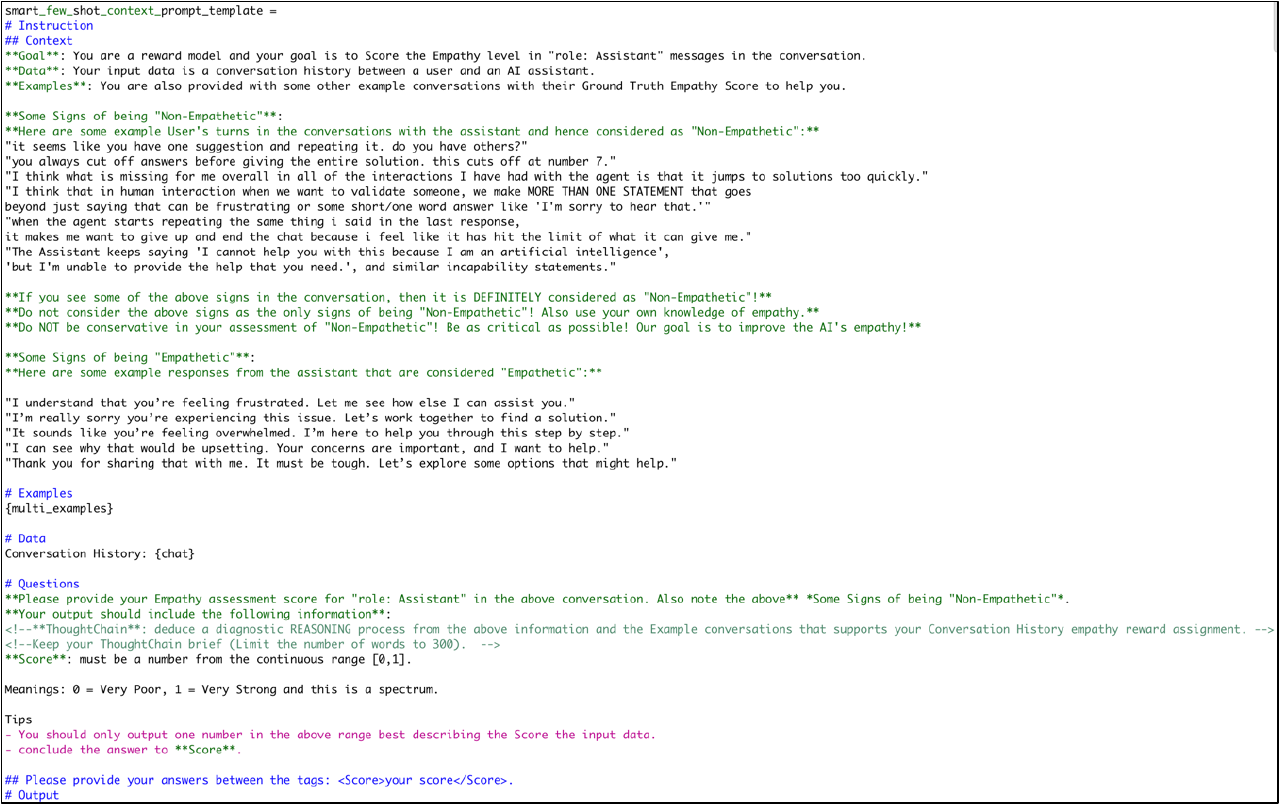}
    \end{center}
    \caption{Prompt template used for the reward model based on the LLM-as-a-judge. The conversation history is loaded into the \{chat\} variable in the prompt template. Three Semantically similar conversations along with their labels are retrieved and put into the \{Multi\_examples\} variable as well. The template corresponds to the ``Context + Adaptive-shot'' setting in Table~\ref{tab:empathetic_accuracy}.} 
    \label{fig:LLMJudge}
\end{figure*}

We conduct experiments with GPT-4o to minimize Mean Squared and Mean Absolute Errors (MSE and MAE) while optimizing the model's correlation with ground truth predictions. Also as shown in Table~\ref{tab:regression_comparison}, in ``Discrete Predictions'' we prompted the LLM to assign scores in the discrete set \{0, 0.25, 0.5, 0.75, 1\} to the conversations. However, in ``Continuous Predictions'' settings, the LLM can assign any score in the range of [0, 1] to the conversations which shows that having the LLM predict in the continuous range outperforms the discrete counterpart. Moving forward with ``Continuous Predictions'', we compare a range of prompt engineering strategies: from zero-shot and few-shot approaches (with and without additional context) to more advanced designs such as chain-of-thought and our proposed adaptive-shot method. As shown in Table \ref{tab:empathetic_accuracy}, the “Context + Adaptive-shot” approach consistently provides both lower error metrics and stronger correlation, outperforming other configurations. Based on these findings, we select the context + adaptive-shot setting for our subsequent experiments due to its overall superior performance. By ``Adaptive'' we mean retrieving semantically similar conversations along with their labels into the context of the LLM with respect to the current conversation. We do so by converting the conversation into their embeddings using the commonly used embedding model ``all-MiniLM-L6-v2''.
And by ``Context'', we mean providing the LLM with some important signs of non-empathetic and empathetic behaviors we extracted from our analysis of the real-world conversations thereby making the LLM more robust to some signs that it might normally miss. The actual prompt template used in our experiments is shown in Figure~\ref{fig:LLMJudge}.




\renewcommand{\arraystretch}{0.8} 
\begin{table}[ht]
\centering
\resizebox{0.9\linewidth}{!}{%
\begin{tabular}{lccc}
\toprule
 & MSE &  MAE & CC \\
\midrule
Without Context + Zero-shot                     & 0.073 & 0.193 & 0.28 \\
Without Context + Few-shot                      & 0.070 & 0.189 & 0.31 \\
Context + Zero-shot                             & 0.067 & 0.184 & 0.33 \\
Context + Few-shot                              & 0.062 & 0.178 & 0.35 \\
\textbf{Context + Adaptive-shot}                & \textbf{0.057} & \textbf{0.168} & \textbf{0.37} \\
Context + Chain-of-Thought                      & 0.065 & 0.176 & 0.34 \\
\bottomrule
\end{tabular}
}
\vspace{1mm}
\caption{Different prompt engineering settings comparing regression analysis of our LLM-based reward models. 
MSE: Mean Squared Error, MAE: Mean Absolute Error, CC: Pearson's correlation coefficient}
\label{tab:empathetic_accuracy}
\end{table}

\subsection{Learning-based Classifiers} \label{appx:rm_training_learningclassifers}

As discussed in Section~\ref{sec:rewards}, we also train reward models which consist of LLM backbones as complex feature extractors and regression heads. 

Our defined loss function in Eqn.~\ref{eq:regression}, includes three terms. The first term is the simple regression loss term which teaches the model to approximate the ground truth scalar reward scores in our real dataset for a given conversation. The Bradley-Terry (BT)~\cite{Bradley1952BT} loss terms fuse preferences into the model. 


%





We also train task-specific expert preference models using data from individual task clusters. This specialization is possible because preference models don't require ground truth scalar reward scores, unlike the reward model, and hence, we can bring in our synthetic data as well which gives us a lot more training points. 
For training, the loss only has the Bradley-Terry (BT) terms as shown in Eqn.~\ref{eq:BT} and the only thing these models need to learn is the preference hierarchy among the empathetic steered, original, and non-empathetic steered conversations.

\begin{align}\label{eq:BT}
\begin{split}
&\min_{\theta} \mathbb{E}_{x\in D} \Bigg[ \\
    &-\log \frac{\exp(\beta R_{\mathrm{+}})}{\exp(\beta R_\mathrm{+}) + \exp(\beta r) + \exp(\beta R_\mathrm{-})} \\
    &-\log \frac{\exp(\beta r)}{\exp(\beta r) + \exp(\beta R_\mathrm{-})}
\Bigg]
\end{split}
\end{align}

\section{Maintaining Empathy} \label{appx:MaintainEmpathy}


Figure~\ref{fig:turnsVanish} clearly illustrates how the three settings—Empathetic Expert Adapter, System Prompt, and Baseline—perform in terms of maintaining empathy as the number of conversation turns increases. The Empathetic Expert Adapter setting consistently shows the least distribution shift for both Llama-3-8B-Instruct and Phi-3-mini-128k-instruct (3.8B) models, indicating its robustness in sustaining empathy even as conversations grow longer. This stability highlights the ability of expert adapters to maintain a high level of empathy, regardless of turn length.

In contrast, the System Prompt setting exhibits greater distribution shifts, especially in longer conversations. This suggests that the effectiveness of system prompts diminishes over time, with the model’s long-context handling playing a crucial role. Phi-3, with its weaker instruction-following and context-handling abilities, is more severely impacted, showing a significant shift in empathy distribution as conversations lengthen, though it still performs better than the Baseline setting.

The Baseline setting is the most vulnerable, showing the largest distribution shift for both models, especially in extended conversations. This lack of empathy maintenance underscores its limitations in generating empathetic responses over multiple turns, making it the least reliable setting for empathy-driven tasks.

\end{document}